\documentclass[sigconf]{acmart}

\AtBeginDocument{%
  }

\copyrightyear{2023}
\acmYear{2023}
\setcopyright{rightsretained}
\acmConference[CHI '23]{Proceedings of the 2023 CHI Conference on Human Factors in Computing Systems}{April 23--28, 2023}{Hamburg, Germany}
\acmBooktitle{Proceedings of the 2023 CHI Conference on Human Factors in Computing Systems (CHI '23), April 23--28, 2023, Hamburg, Germany}\acmDOI{10.1145/3544548.3581371}
\acmISBN{978-1-4503-9421-5/23/04}

\usepackage{color}
\usepackage{textcomp}
\usepackage{multirow}
\usepackage{subcaption}
\usepackage{bm}
\usepackage{wrapfig}
\usepackage{xspace}
\usepackage{longtable}
\usepackage{algorithm}
\usepackage{algpseudocode}
\algtext*{EndFor}
\algtext*{EndIf}

\renewcommand{\shortauthors}{Kang et al.}

\newcommand{\xhdr}[1]{\vspace{1mm}\noindent{{\bf #1.}}} 
\newcommand*\cirnum[1]{\raisebox{.5pt}{\textcircled{\raisebox{-.9pt} {#1}}}}
\newcommand{\sys}{\textsc{ComLittee}\xspace}
\newcommand{\system}{\sys}
\newcommand{\baseline}{\textsc{Baseline}\xspace}

\newcommand{\generalspace}{author-augmented literature discovery\xspace}
\newcommand{\eg}{\textit{e.g., }}
\newcommand{\cf}{\textit{cf. }}

\newcommand{\ie}{\textit{i.e., }}
\newcommand{\Ie}{\textit{I.e., }}
\newcommand{\tind}[3]{$t_{\text{two-tailed}}$(#1)=#2, $p$=#3}
\newcommand{\tpaired}[3]{$t_{\text{paired}}$(#1)=#2, $p$=#3}
\newcommand{\chisq}[2]{$\chi^2(1)$=#1, $p$=#2}
\newcommand{\mannw}[2]{\text{Mann-Whitney} $U$=#1, $p$=#2}
\newcommand{\wilcoxon}[2]{$\text{Wilcoxon}$ $W$=#1, $p$=#2 }

\newcommand{\todo}[1]{{\textcolor{red}{ }}}

\newcommand{\bug}
    {\mbox{\rule{2mm}{2mm}}}

\newcommand{\jb}[1]{\textcolor{blue}{\bug \footnote{\textcolor{blue}{ }}}}
\newcommand{\jbin}[1]{\textcolor{blue}{ }}
\newcommand{\jc}[1]{\textcolor{brown}{}}
\renewcommand{\jc}[1]{}

\newcommand{\new}[1]{{#1}}
\newcommand{\nnew}[1]{{#1}}

\begin{document}

\title[\sys]{\sys: Literature Discovery with\\
Personal Elected Author Committees}

\author{Hyeonsu B. Kang}
\authornote{Work completed during a research internship at Semantic Scholar Research, Allen Institute for AI.}
\orcid{0000-0002-1990-2050}
\affiliation{%
  \institution{Carnegie Mellon University}
  \streetaddress{5000 Forbes Ave}
  \city{Pittsburgh}
  \state{PA}
  \country{USA}
}
\email{hyeonsuk@cs.cmu.edu}

\author{Nouran Soliman}
\orcid{0000-0001-8846-2098 }
\affiliation{%
  \institution{MIT CSAIL}
  \streetaddress{32 Vassar St}
  \city{Cambridge}
  \state{MA}
  \country{USA}
}
\email{nouran@mit.edu}

\author{Matt Latzke}
\affiliation{%
  \institution{Allen Institute for AI}
  \streetaddress{2157 N Northlake Way \#110}
  \city{Seattle}
  \state{WA}
  \postcode{98103}
  \country{USA}
}
\email{mattl@allenai.org}

\author{Joseph Chee Chang}
\orcid{0000-0002-0798-4351}
\affiliation{%
  \institution{Allen Institute for AI}
  \streetaddress{2157 N Northlake Way \#110}
  \city{Seattle}
  \state{WA}
  \postcode{98103}
  \country{USA}
}
\email{josephc@allenai.org}

\author{Jonathan Bragg}
\orcid{0000-0001-5460-9047}
\affiliation{%
  \institution{Allen Institute for AI}
  \streetaddress{2157 N Northlake Way \#110}
  \city{Seattle}
  \state{WA}
  \postcode{98103}
  \country{USA}
}
\email{jbragg@allenai.org}

\begin{abstract}
In order to help scholars understand and follow a research topic, significant research has been devoted to creating systems that help scholars discover relevant papers and authors.
Recent approaches have shown the usefulness of highlighting relevant authors while scholars engage in paper discovery. However, these systems do not capture and utilize users' evolving knowledge of authors.
We reflect on the design space and introduce \sys, a literature discovery system that supports author-centric exploration.
\new{In contrast to paper-centric interaction in prior systems, \sys's author-centric interaction supports curating research threads from individual authors}, finding new \new{authors} and \new{papers} using combined signals from a paper recommender and the curated authors' authorship graphs, and understanding them in the context of those signals.
In a within-subjects experiment that compares to a \nnew{paper-centric discovery system with author-highlighting, we demonstrate how \sys improves author and paper discovery.}

\end{abstract}

\renewcommand{\shortauthors}{H. B. Kang, N. Soliman, M. Latzke, J. C. Chang, and J. Bragg}



\maketitle

\section{Introduction}
\begin{figure*}[t]
    \includegraphics[width=.9\textwidth]{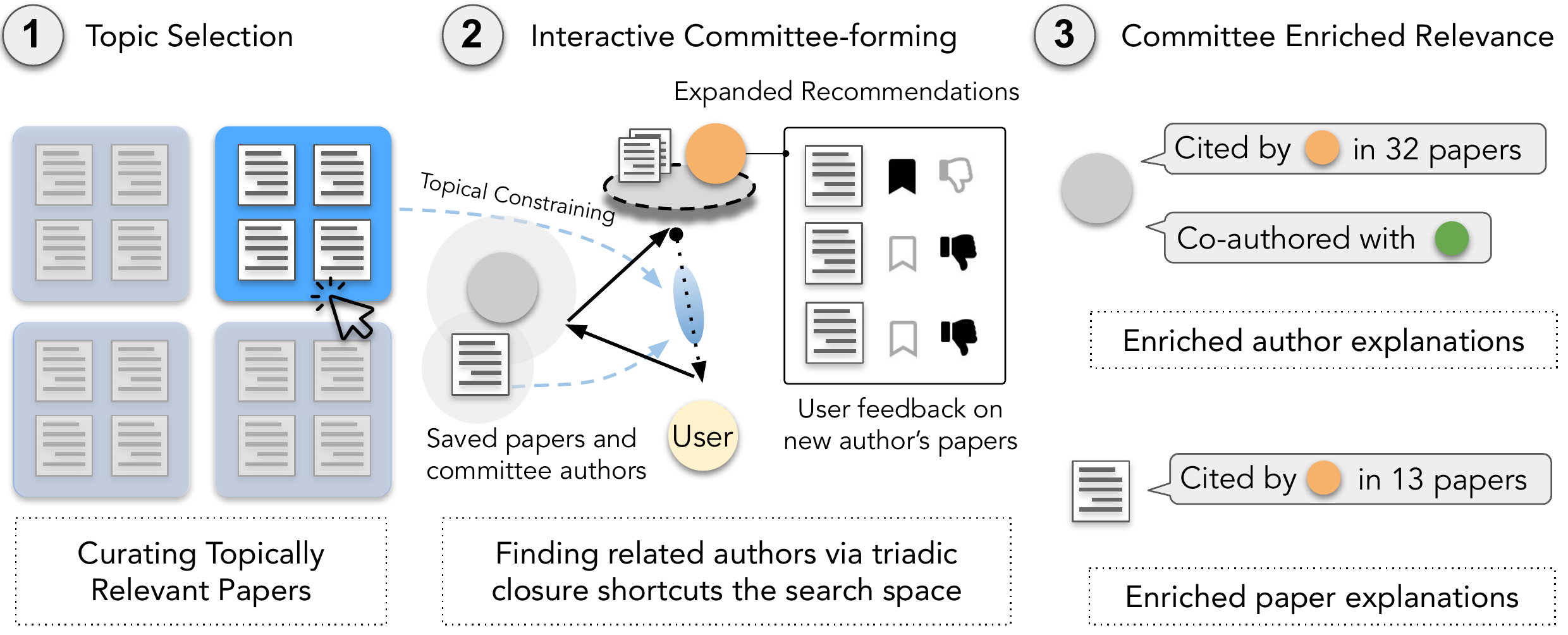}
    \vspace{-1em}
    \caption{Construction and application of relevance signals on \sys: \cirnum{1} The user saves seed papers relevant to a topic. \cirnum{2} \sys recommends an initial set of authors and papers relevant to the selected folder's content. When the user starts forming a committee of authors, \sys expands the recommended set of authors and papers using the committee authors' citation network as well as by sourcing authors who published similar papers, based on a prediction of relevance analogous to the phenomenon of triadic closure~\cite{granovetter1973strength} observed in social networks. During the expansion, topical relevance represented as a set of user feedback on papers is used to constrain the expansion. \cirnum{3} The curated and maintained committee is applied to enrich relevance signals in subsequent author and paper recommendations. In the figure, no-index circles represent authors.}
    \label{fig:main}
    \Description{A diagrammatic of system workflow on ComLittee is shown. The first panel describes how users curate a topic folder and selects it to start receiving recommendations. The second panel shows how ComLittee uses users' positive or negative feedback on a recommended author and her papers to source new author recommendations. The third panel shows how the relevance explanation is featured.}
\end{figure*}
In order to understand a research area and keep up with the exponentially growing rate of scientific publication (\cf~\cite{bornmann2021growth, jinha2010article, van2014global}), scientists expend significant effort searching for relevant papers and threads of research. \new{To help scientists in different stages of the discovery process many systems have been developed to support finding, triaging, and reading papers (\cf~\cite{citeread,litsense,paperquest,choe2021papers101,CiteSense,wang2016guided}). While useful, such systems often limit interaction to the modality of papers and do not incorporate useful relevance signals from authorship graphs.} \nnew{However,} identifying authors that work in a research area can further facilitate the discovery process, since authors often publish multiple papers on related topics as they pursue a research agenda, which can be discovered together. Further, the publication networks of these authors could also help with identifying additional relevant authors and papers, for example, through their frequent co-authors who also work on the topic, the relevant papers they often cite, and authors of these relevant papers.

While much research has been devoted to creating literature discovery tools, given these benefits of identifying relevant authors, recent research has begun to develop a new 
kind of discovery tool that augments encountered papers with highlights of potentially-relevant authors on those papers \cite{kaur-uist22-feedlens,chi22_from_who_you_know}.
Users can use this information to help them decide whether to read encountered papers~\cite{chi22_from_who_you_know}, or explore additional papers written by highlighted authors, which may be relevant~\cite{kaur-uist22-feedlens}.
These highlights have been demonstrated to be effective in multiple literature discovery contexts including exploratory paper search~\cite{kaur-uist22-feedlens} and paper recommendation~\cite{chi22_from_who_you_know}.
We term this emerging paradigm of literature discovery systems that 
support author discovery~\textit{\generalspace}.

While highlighting relevant authors provides useful context for users of literature discovery systems, current systems give users limited ability to interact with authors.
First, because these systems augment encountered papers but do not alter the underlying methods for recommending papers, they have limited ability to help users explore and discover authors beyond those who have authored the set of surfaced papers.
Users may miss important threads of research because the returned papers are not optimized to yield good coverage of relevant authors; these systems typically return papers prioritized by predicted relevance, from a sample of a larger corpus, and may fail to target a group of authors that builds on each other's work, \nnew{which would} provide valuable context \nnew{about} research contributions. 
Second, these systems do not let users save authors they have discovered and know to be relevant, which limits the system's ability to help users make connections to new papers and authors through these known authors over time.
\nnew{Without saving functionality}, systems can only recommend authors that are predicted to be relevant. 
And because users may have limited interest in or knowledge of these recommended authors, \nnew{they may be confused or discouraged}  from using the recommended authors as rich mechanisms for discovering and explaining the relevance of additional papers and authors~\cite{chi22_from_who_you_know}.

Here we propose \sys (\autoref{fig:main}), an \generalspace system that promotes rich author-centric interaction.
In order to inform the design of \sys, we formulated a design space for \generalspace\ (Table~\ref{tab:design-space}), in which \sys\ covers new ground along several dimensions.
In terms of the workflow, users on \sys can 1) select a topic and initialize the system with personally curated seed papers; 2) receive author recommendations produced from a larger set of recommendation sources compared to prior work, using a new approach that combines paper recommender scores and publication network relations with previously-saved authors; 3) save (elect) relevant authors over time to a personal committee, which iteratively updates the system to inform future exploration and enable rich explanations of the relevance of both the recommended author and the papers contained in the recommended author card. In addition, \sys renders the relevance explanations as interactive filters useful for quickly homing in on specific papers that contribute to the relevance between authors. \nnew{We instantiate these interaction and the author-centric workflow features as a list-based discovery system, enabling comparison to related systems~\cite{kaur-uist22-feedlens,chi22_from_who_you_know}, and evaluate it in a controlled laboratory study to uncover their feasibility, value, and implications for future design.}





In summary, this work makes the following contributions:
\begin{itemize}
    \item We present a design space for \generalspace,
    with
    \nnew{seven} interaction and presentation primitives, to situate the current work within the literature and to inform future designs in this
    emerging space.
    \item We propose \sys, a novel interactive author-centric system for \generalspace. 
    \item We evaluate \sys in a within-subjects study ($N=16$), and demonstrate its \new{value} over a 
    \nnew{strong paper-centric} baseline based on a prior system~\cite{kaur-uist22-feedlens}. \nnew{Through} detailed quantitative, behavioral, and qualitative analyses, \nnew{we report how \sys led to gains in discovery efficiency (for both authors and papers), novelty (for authors), and interestingness (for papers). In addition, we} provide implications for design for future systems in \generalspace.
\end{itemize}

\section{Related Work}
\begin{table*}[h!]
    \centering
    \caption{The design space for \generalspace. We situate \sys and the closest prior work in terms of \new{seven design axes} as follows \new{(see text for descriptions). 
    }
    \new{Who:} Curated or Inferred.
    \new{What}: Paper (P) or Author (A).
    \new{When: Sequential (S) or Batched (B) effects.}
    \new{Where}: Push or Highlight.
    \new{Why}: Explanation (Exp) or Recommendation (Rec).
    \new{How (Relevance Source)}: Paper recommender score (R); Co-author relation (Co); Cited author relation (Ci); Prior user interaction history (I); \new{or Other mechanism (Misc.)}.
    \new{How (Relevance Distance)}: Direct (D) or Indirect (I).
    }
    \vspace{-1em}
    \begin{tabular}{c|cccccccc}
      \toprule
      &
      \begin{tabular}{@{}c@{}}\new{Who} \end{tabular} &
      \begin{tabular}{@{}c@{}}\new{What} \end{tabular} &
      \begin{tabular}{@{}c@{}}\new{When} \end{tabular} &
      \begin{tabular}{@{}c@{}}\new{Where} \end{tabular} &
      \begin{tabular}{@{}c@{}}\new{Why} \end{tabular} &
      \begin{tabular}{@{}c@{}}\new{Relevance}\\\new{Source} \end{tabular} &
      \begin{tabular}{@{}c@{}}\new{Relevance}\\\new{Distance} \end{tabular} \\
      \midrule
      \sys & Curated & P,A & S & Push & Exp & R,Co,Ci & I  \\
      FeedLens~\cite{kaur-uist22-feedlens} & Inferred & P,A & S & Highlight & Rec/Exp & R & D \\
      \citet{chi22_from_who_you_know} & Inferred & P & B & Highlight & Rec/Exp & Co,Ci,I & D,I  \\
      \new{Bridger~\cite{bridger-chi22}} & Inferred & A & -- & Push & -- & Misc. & D \\
      Baseline & Curated & P & S & Highlight & Rec/Exp & R & D \\
      \bottomrule
    \end{tabular}
    \label{tab:design-space}
    \Description{The system interface and its various features are shown.}
\end{table*}

\new{Below we review existing paper-centric systems (Section~\ref{subsection:paper-centric}); how theoretical and empirical studies of expertise-finding systems suggest the importance of author modality in the literature discovery process (Section~\ref{subseciton:expert-finding});
selected work from interactive machine learning, which guides our research questions (Section~\ref{subsection:iml});
and various examples of relevance explanation, which guide our system design (Section~\ref{subsection:relevance_explanation}).}
\subsection{Paper-centric Literature Discovery Systems} 
\label{subsection:paper-centric} \new{To help scientists \nnew{and professionals~\cite{mysore2023data}} in various stages of the literature discovery process, significant research effort has been devoted to developing interactive systems. However, most prior systems have focused on \emph{documents} (i.e., research papers) as the core primitive for both user interactions and discovery, and do not support exploring social signals (i.e., authors and their \nnew{relations to other authors}) around the documents.
For example, {Papers101}~\cite{choe2021papers101} helps scholars search for additional relevant literature by generating unused keywords for query expansion. Kang et al. showed that enabling search for analogous papers based on the purpose-mechanism schema~\cite{kang_augmenting_tochi} or diverse domains~\cite{naacl2022_kang_augmenting} can increase creativity of scientists' ideas. Once a set of papers has been discovered, systems can support subsequent tasks; \eg{PaperQuest}~\cite{paperquest} with triaging which papers to read next, {LitSense}~\cite{litsense} with overviews and filtering of a collection of searched papers, {CiteSense}~\cite{CiteSense} with appraisal and grouping papers, {Threddy}~\cite{threddy} with organizing papers into notable threads of research while reading, and Wang et al.'s system~\cite{wang2016guided} with visualizing collected papers in a broader narrative structure. 
In addition, {Passages}~\cite{han2022passages} helps users collect text snippets while reading papers, which can be re-represented into a relational form (\eg matrix) later. \nnew{Relatedly~\cite{relatedly} helps users discover relevant paragraphs from papers, and CoNotate~\cite{conotate} and Interweave~\cite{interweave} help by expanding queries that promote `active' searching~\cite{active_search}}. \nnew{Finally, several systems have been developed to help reduce the cognitive cost of reading papers and documents (\eg ScholarPhi~\cite{scholarphi}, CiteRead~\cite{citeread}, Scim~\cite{citeread}, Fuse~\cite{fuse}, Crystalline~\cite{crystalline}, and Wigglite~\cite{wigglite}).} Despite differences in use case scenarios, these systems share a commonality in design that centers \textit{papers} as the \nnew{mode} of interaction. In \sys, we also focus on designing an interactive system that can help users explore and discover papers relevant to a topic of interest. In contrast to the above prior work, we treat \emph{authors} as the \nnew{mode} of interaction and discovery, while at the same time allowing participants to actively explore relevant papers in the context of relevant authors.}

\subsection{Expert-finding Systems} \label{subseciton:expert-finding} \new{A parallel line of research on expert and social recommendation has also been developed (see~\cite{tang2013social,guy2011social} for a review).} Systems such as ReferralWeb~\cite{referralweb} \new{and Expertise Recommender~\cite{expertise_recommender} contributed understanding} of \new{social conditions in which} expert recommendations \new{become useful, while} Guy et al.~\cite{guy2013mining} found microblogs as a valuable source of data for expertise-matching. A particularly relevant line of research \new{explored authorship networks}, such as \new{Liben-Nowell and Kleinberg's work}~\cite{liben2003link} on predicting future collaborations among scholars using co-authorship networks, and Conference Closure~\cite{from_triadic_closure_to_conference_closure} which proposed a new form of triadic closure (\ie scholars who attend mutual conferences are more likely to form connections) and studied its effects on future collaborations. \new{As a specific form of expert recommendation,} Bridger~\cite{bridger-chi22} proposed algorithms that facilitate \new{\textit{author} recommendations and burst filter bubbles by} allowing users to select facets \new{of expertise} from an author's prior publications\new{, and matching on the selected factors while diversifying on others in recommendation}. \new{Theoretical work such as Burt's `Structural Hole' theory~\cite{burt2004structural} suggests the importance of author recommendation for brokering knowledge across different fields, and empirical evidence suggests an increasing importance in the face of deepening specialization of knowledge~\cite{rzhetsky2015choosing,swanson_undiscovered}. Studies that examined the practice of literature discovery, such as Sandstorm's~\cite{sandstrom2001scholarly} and Pirolli's~\cite{pirolli2009elementary}, showed how authors play a highly valuable role in the process. \nnew{These studies} also give inspirations to recent work} by Kang et al.~\cite{chi22_from_who_you_know} and Kaur et al.~\cite{kaur-uist22-feedlens} that leveraged authorship graphs to augment encountered papers with highlights of potentially-relevant authors in them\new{, showing improved user engagement and discovery experience. Taken together, author recommendation systems and recent work that leverages authorship graphs suggest a burgeoning design space for future \generalspace systems; hence, we use them to form the bases of our design space (Section~\ref{sec:design-space}).}

\subsection{Interactive Machine Learning} \label{subsection:iml}
\new{Though literature discovery can be modeled as users submitting a series of independent search queries and reviewing the retrieved results for each, complex user intent and evolving knowledge may be better served by a system that can interactively learn from user feedback throughout the course of discovery. Here, the field of interactive machine learning offers relevant examples and insights.} A particularly relevant example is the Regroup system~\cite{regroup}, which proposed a novel probabilistic approach that iteratively updates priors based on a user's feedback on group members as they curate them. \new{A core insight from this work is that search-by-name and system-generated recommendations have complementary strengths, with the former effective for forming small, well-defined groups while the latter helpful for large, varied groups. Another insight comes from Kocielnik et al.'s study~\cite{kocielnik2019willyou} that showed how a recall-oriented machine learning system objective improves user perception and willingness to adopt over a precision-oriented objective at equal performance levels. This is also consistent with how users benefited from diverse author recommendations on Bridger~\cite{bridger-chi22}. The question is then, how can users effectively and continuously discover novel authors?
Addressing this question requires studies that analyze user and system behavior over time to shed light on effective user navigation strategies in the context of changing alignment between human- and AI-model of relevance. Our study and collection of behavioral data in the course of user interaction was designed to contribute to this gap in the literature.}
\subsection{Relevance Explanation} \label{subsection:relevance_explanation}
\new{
One key component of interactive recommender systems is supporting users in making sense of the recommended items. For example, SearchLens~\cite{chang2019searchlens} and FeedLens~\cite{kaur-uist22-feedlens} adopted a \emph{lens} metaphor and provided interactive at-a-glance explanations and relevance filters. Findings in RelevanceTurner~\cite{relevance_turner} also showed the benefits of making recommendation sources more transparent in discovery tasks. One promising approach in prior work is to explain newly recommended items by drawing connections to previously discovered or familiar items. For example, Apolo used a \emph{relative spatial layouts} design where users can iteratively explore parts of the citation graph around familiar papers~\cite{apolo}, and Kang et al. showed benefits in providing personalized relevance explanations around emailed paper recommendations, \nnew{which} described connections between the user and the authors of the papers~\cite{chi22_from_who_you_know}.
When designing \sys, we were inspired by the high-level ideas in the above prior work. Specifically, we adopted the ``lens'' metaphor at both the author and paper levels to surface ones that were most relevant. In addition, we also drew connections between a recommended author to the set of authors already familiar and saved by a user as a way to explain their relevance.
}

\begin{figure*}[t]
    \includegraphics[width=.95\textwidth]{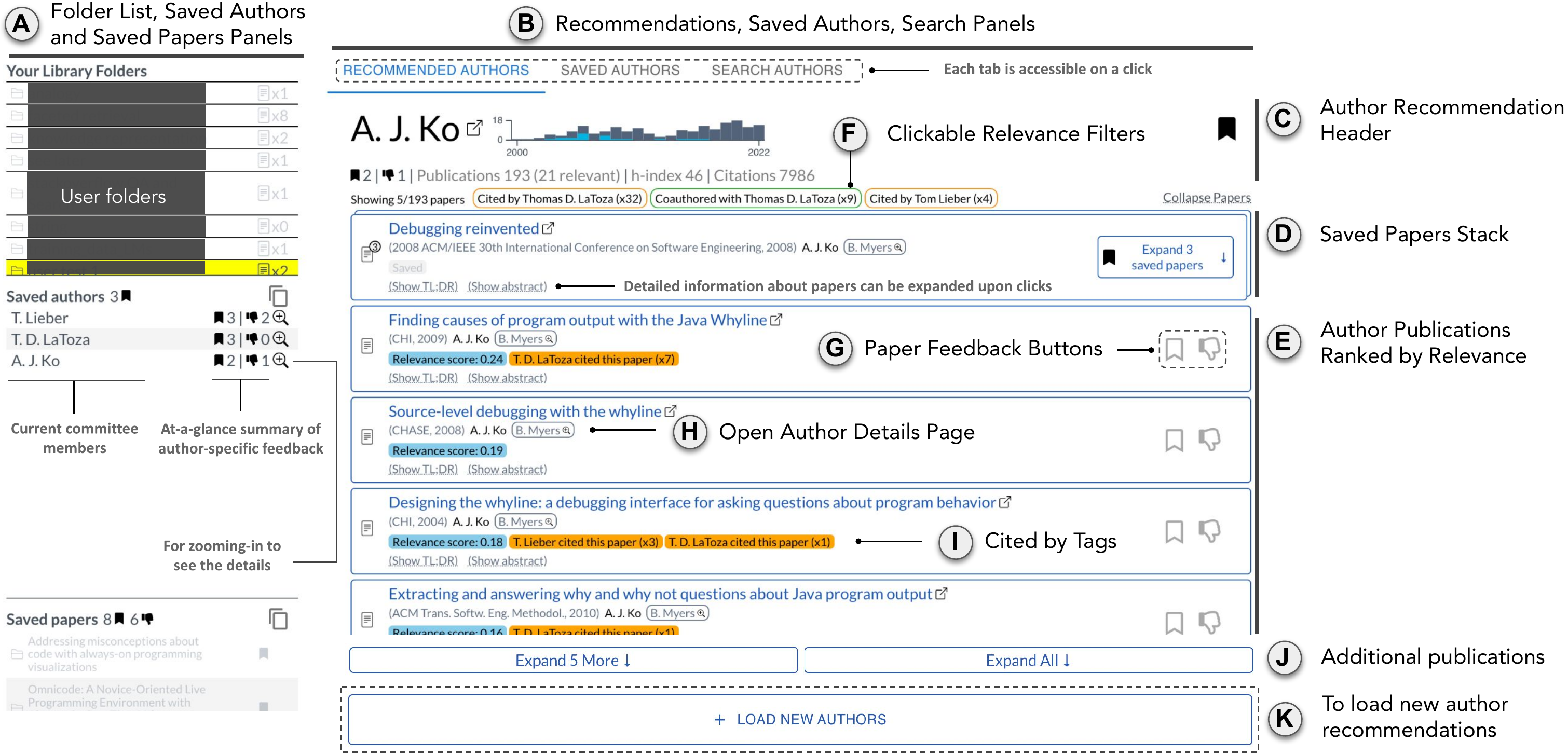}
    \vspace{-1em}
    \caption{\sys features: \cirnum{A} topic folders, saved papers and authors (\ie currently saved committee members) panels, \cirnum{B} \new{the primary} recommendations \new{tab} and direct \new{search-by-name} of authors (not shown), \cirnum{C} -- \cirnum{J} various features \new{for interacting with} author recommendations (see text), \cirnum{K} button for generating new recommendations \new{with updated user feedback}.
    }
    \label{fig:interface}
    \Description{Main ComLittee interface consisting of a side bar on the left and a list interface with author recommendations on the right.}
\vspace*{-8pt}
\end{figure*}

\section{Design space for \generalspace}
\label{sec:design-space}
Recently, three prior works on literature discovery systems that incorporate author entities (described in Section~\ref{subseciton:expert-finding})~\cite{kaur-uist22-feedlens,chi22_from_who_you_know,bridger-chi22} have also explored various designs for \generalspace. In order to situate our work and inform future work, we surveyed \nnew{past design decisions} and formulated a design space (\autoref{tab:design-space}) for \generalspace. \new{The prior work was contrasted to \sys and the baseline implementation, for contextualizing our evaluation and convenience of reference. The seven axes represent dimensions ranging from the scope of user controllability to how author recommendations were sourced.
The axes were chosen not only for their coverage of salient features of system designs in the space,
but also for their generative potential; though we demarcate categorical values along each axis in the space here, future work may use it to envision novel system designs that either instantiate new values along a specific axis, expand the space by contributing new axes, or by proposing new combinations of values along the axes.}


\begin{itemize}
\item \textit{\new{Who (User controllability and agency in iterative
steering of author recommendations:} Curated vs. Inferred).} \new{One of the salient design choices lies in the decision of how much user controllability the system supports in iterative steering of author recommendations, and conversely how much should be automated. On one end of the spectrum is designs without any direct curation support, which sacrifices users' steering capabilities in favor of the convenience of minimum required effort. Previous work}~\cite{chi22_from_who_you_know} \new{instantiated this design choice by} inferring reference authors thought to be of interest (Indirect reference authors) \new{from user publication and interaction logs,} which were then used as proxies for \new{the user to help identify relevant authors}. \new{In contrast,} \sys supports \new{manual curation by directly adding/removing} indirect reference authors \new{to/from an elected committee}.

\item \textit{\new{What (Explained Entity Types:} Paper vs. 
Author).} \new{The information algorithmically gleaned from how relevant authors were found (e.g., a paper of theirs was cited by a relevant author)} may be featured as explanations on an author's papers (\eg when viewing on author page with publications), or in an aggregate form as an author summary\new{, which was shown to improve user understanding and engagement~\cite{chi22_from_who_you_know}}. FeedLens~\cite{kaur-uist22-feedlens} provides paper- and author-level explanations using recommender scores. \sys\ \new{extends this to incorporate explanations} based on citation and coauthor relations, which show the relation of the papers to authors of interest.

\item \textit{\new{When (Delay between user feedback and system
changes: Sequential  vs. Batched).}} \new{Another salient design choice lies in how quickly the system responds to user feedback. When steering author recommendations early on, users may need to see effects take place immediately upon their most recent feedback to form a mental model of the system, requiring sequential (immediate) rather than batched (slow) updates. However, enabling low-latency sequential updates may require sacrificing some degree of accuracy~\cite{kaur-uist22-feedlens}. 
Furthermore, whether and how divergence between the most recent snapshot of user feedback and the overall feedback the system has accumulated over time may manifest in the literature discovery process remains an open empirical question.
}

\item \textit{\new{Where} \new{(Whether authors are presented as top-level push} recommendations vs. In-situ highlights \new{within an encountered paper context}).} A system can either `push' authors as top-level recommendations, or highlight them {\it in situ} in the context of papers the user encounters. \sys\ and author recommenders (\eg Bridger~\cite{bridger-chi22}) do the former. Most literature discovery systems do neither, but recently some do the latter by highlighting authors in the context of papers the user encounters during exploratory search~\cite{kaur-uist22-feedlens} or paper recommendations~\cite{chi22_from_who_you_know}.

\item \textit{\new{Why (The purpose of author highlights: explanatory} vs. \new{to recommend novel authors}).} In-situ author highlights can either make you aware of authors you know, or highlight new authors you may wish to explore. The former can be used as an explanation of the relevance of a paper to you~\cite{chi22_from_who_you_know}, whereas the latter is more useful \new{as a launchpad for} exploration~\cite{kaur-uist22-feedlens}. Systems may not know which authors the user knows, so highlights may function as either type depending on user knowledge. \sys\ only shows in-situ explanations, so as not to confuse users by recommending authors in-situ and also as top-level entities.

\item \textit{\new{How: Relevance Source} via recommender (R) vs. author (Co, Ci) vs. user (I).} Author relevance may be determined by scoring relevance of their papers using a paper recommender algorithm~\cite{kaur-uist22-feedlens}, by identifying their citation (Ci) or co-author relations (Co) to other (reference) authors of interest~\cite{chi22_from_who_you_know}, or by identifying authors based on the user's interaction history (I) such as how many times the user has saved their papers to their library~\cite{chi22_from_who_you_know}. \sys\ determines relevance using all three of these categories, and additionally explores leveraging a new aspect of the user's interaction history---the authors they have saved to their committee. \new{In contrast, Bridger~\cite{bridger-chi22} makes author recommendations using faceted content mined from the text of each author's publications (denoted as Misc. in \autoref{tab:design-space}) rather than traversal on authorship and citation graphs.}

\item \textit{\new{How: Relevance Distance via} Direct vs. Indirect reference author.} The reference authors of interest from which relevant co-authors and cited authors are determined (see \new{Relevance Source} above) may either be the user (if they are an author), or a different author of interest. \citet{chi22_from_who_you_know} terms the former ``Direct'' relations and the latter ``Indirect'' relations (expressing the further, indirect relation to the user in that case, via a proxy author of interest). Most author recommendation systems like Bridger~\new{\cite{bridger-chi22}} focus on direct relations to the user. \citet{chi22_from_who_you_know} explore both; however, indirect authors are inferred, which can be confusing to the user when they are not relevant or the user does not know about them. \sys\ overcomes these issues by allowing users to specify indirect authors to the system. \sys\ does not currently implement direct author relations since our experimental task is focused on finding entities related to a topic rather than the user in general.
\end{itemize}

\section{System Description} \label{section:system}
\system uses \new{various sources of} signals to identify and provide both author and paper recommendations. To start, we assume user\new{s can} find a small set (i.e., $\geq 5
$) of papers relevant to a topic to initialize the underlying paper recommender (detailed below). \new{Though not strictly required,} we also assume users might already know \new{some} authors relevant to the topic \new{that may constitute their} initial ``committee.'' \new{The system then} help\new{s} users discover other relevant authors and their papers, \new{while surfacing} connections between familiar \new{committee} authors and new authors.
\subsection{Example User Scenario}
A research scientist \new{wants to} learn more about a new research area that she recently started \new{exploring}. She uses a scholarly search engine and an online paper recommender service to discover recent papers based on a few papers that she had saved for the topic. However, she was also interested in learning more about who the relevant authors in the field were, which would allow her to reach out to them for potential collaborations. From reading papers, she notice\new{s} a few with relevant recent \new{publications} that she liked, but neither a paper recommender nor search engines can support her to proactively explore more relevant authors.

Feeling frustrated she switches to \system and imports the relevant papers in her library folder to the system. She \new{can} immediately see a list of authors relevant to the topic, some of whom were familiar to her
from recent reading.
Recognizing \new{them}, she quickly click\new{s} the bookmark buttons next to their names (Fig.~\ref{fig:interface}c) to save them to her ``committee'' for the topic. In addition, she also learn\new{s} that some of the familiar authors had written relevant papers she did not know about, which were ranked to the top of their publications page based on the relevance to the folder (Fig.~\ref{fig:interface}e). Using the Paper Feedback Buttons (Fig.~\ref{fig:interface}g) she \new{can} save additional relevant papers to her folder, or downvote irrelevant papers to help {\system} better understand her interests.

\new{Later} she click\new{s} on the Load New Authors button (Fig.~\ref{fig:interface}k) to get a new batch of recommendations. Now she notices \sys starting to recommend authors that she does not recognize. In their author cards, she \new{can} see their connections to authors she already knew. For example, one unfamiliar author was frequently cited by several familiar authors, and she could see multiple interesting and relevant publications in the paper list. Feeling confident, she save\new{s} this new author to her folder. As she saves more, \sys also continues to find more \new{relevant} authors, improves its explanations about them based on \new{committee} authors, and ranks their papers with better accuracy. 
\subsection{Author-Centric  Recommendation Strategies}
To support the features described above, \sys introduces the following four strategies that \emph{expands} from papers and authors already familiar to the users to recommend \new{unknown} authors and papers for discovery.
\subsubsection{Library-extracted} 
The user may start using \system with only few saved papers in a library folder. The user may be familiar with some of the papers, their concepts, and/or authors, which prior work showed could be effective for boosting user engagement in the email paper recommendation alert context~\cite{chi22_from_who_you_know}.

\xhdr{Procedure} Given a few saved papers, authors \new{of} the papers are tallied and sorted in a descending order of frequency.

\subsubsection{Authored multiple relevant new papers}
Based on papers saved in the user's folder or papers they had downvoted (Fig.~\ref{fig:interface}g), \sys searches an index of recent publications for 100 papers that were the most topically relevant (within the last 180 days, as commonly used in public paper recommenders). These 100 papers then cast 100 `votes' on their authors. 
The votes on authors are tallied and sorted in a descending order of counts.

\xhdr{Implementation of the relevance prediction model} Similar \new{to}~\cite{kaur-uist22-feedlens}, we use an ensemble regression model that scores each paper on a scale of [-1, 1], where a negative score represents predicted irrelevance and a positive score represents predicted relevance (stronger towards both ends of the scale). The model averages the outputs of two linear Support Vector Machines (SVMs) that differ in terms of its training procedure and specifically how each paper is represented as a feature vector: the first SVM uses textual features (unigrams and bigrams) with Tf-Idf normalization, similar to the public arxiv-sanity recommender (\url{https://arxiv-sanity-lite.com}), and returns high-precision direct matching on terms within search queries. The second SVM uses SPECTER embeddings, with a focus on matching on the multifacted semantic relatedness beyond term-matching of the first SVM, and showed good performance on paper recommendation tasks~\cite{specter}. The ensemble of the two therefore balances the strength of each matching model and is iteratively re-trained in real-time, based on each user's feedback on paper recommendations. \nnew{We used the Semantic Scholar Academic Graph API to access extracted author names from their publications. We refer to~\cite{Subramanian2021S2ANDAB} for documentation on the quality of the name extraction pipeline for interested readers. }


\subsubsection{Coauthorship-based expansion} Users may also find new authors who co-authored with familiar authors they trust for a topic. In such cases, the familiar authors can be viewed as mediating wedge-shape paths between the user and \new{each} new author on the publication \new{graph}.
Recent work~\cite{chi22_from_who_you_know} also showed highlighting author names in paper recommendations based on \new{the same} mechanism of triadic closure~\cite{granovetter1973strength,kossinets2006empirical} on citation \new{graphs} increases user engagement.

\xhdr{Procedure} Starting with user-saved authors, each of the saved author's list of publications is searched and \new{assigned a} relevance score, using the same prediction model described above. When the user has already provided feedback on a paper (\eg when she received one of the coauthors of the paper as a recommended author earlier and encountered the papers), we overwrite the relevance score with either 1 (\ie the user has saved this paper earlier) or -1 (\ie downvoted), in order to treat user feedback as ground truth. Then relevant papers are filtered (\ie have scores $> 0$). \new{Finally,} using a similar voting procedure from papers to their authors as before, authors with the highest votes are collected. 

\subsubsection{Citation-based expansion} Another way trust can be propagated between familiar and unfamiliar authors is through citations in their papers. The assumption here is scholars cite papers they trust in their own papers, so that users may find value in discovering unfamiliar authors through papers frequently cited by trusted and familiar authors.

\xhdr{Procedure} Step 1) Using each user-saved author's publications, we collect up to 100 most relevant papers based on their scores. Step 2) For the collective referenced papers from the papers in Step 1, we design a voting procedure in which each saved author casts 1 vote on a reference if the author has at least one paper that is a) included in the sampled relevant papers in Step 1, and b) cites the reference. We exclude self citations for diversification. We assign votes at the author level rather than the paper level to prevent authors with many publications \new{from} dominating the votes. Step 3) Sample the top 100 references with highest votes (higher vote counts means more of the saved authors have previously cited that work). Step 4) Using the references from Step 3, repeat a similar paper-to-authors voting procedure as earlier, and finally collect the authors with highest votes over the references. \new{See Appendix~\ref{appendix:algorithms} for pseudo-code implementation.}

\subsubsection{Batch generation} \label{section:batch_generation}
Each of the four strategies above generates a ranked list of author recommendations. Initially, the top two recommendations from each list \new{are} selected and interleaved (Appendix~\ref{appendix:sort-authors}) to create a \emph{batch} of eight author recommendations. A cursor is \new{then moved to} point to the next top ranked recommendation \new{in} each list. When users clicks d on the ``Load New Authors'' button (Fig.~\ref{fig:interface}k), \new{a new} batch of recommendations is generated \new{using} the cursors. Whenever the user saves new authors or papers, or down-votes a paper, \new{the lists replenish and cursors are reset to the top}. \sys shows all available explanations for each author, regardless of the strategy it was selected from. For example, if an author was selected because it was frequently cited by familiar authors, but also coauthored with a committee author before, both \new{`cited by' and `coauthored'} explanations are shown. \new{Relevance explanations of} current author recommendations within each batch \new{are} update\new{d} as the user interacts with \sys. 

\subsection{Relevance explanation features} \new{The} author-level explanations \new{above} are translated into interactive relevance explanation filters (Fig.~\ref{fig:interface}f) in \new{each} author recommendation header. \new{An} exception is the information about the number of predicted relevant papers the recommended author published \new{which} is displayed simply as a static text \new{tag} next to the number of total publications (a filter is not needed because papers are sorted by relevance scores by default). Relevance filters also interact with \new{a small} publication year-count histogram next to each author's name; \new{clicked filter} add\new{s} overlays that correspond to the count of papers included in the relevance relation (\eg A ``Coauthored with... (x9)'' filter will bin the 9 papers into published years and add \new{corresponding} visual marks -- bars -- to the vis upon a click). \sys also features paper-level relevance explanations which show the predicted relevance score for each paper at the time of recommendation, and up to three authors who have cited the paper most often, \new{with} the number of their papers that cited \new{it}, \new{while} excluding self citations.

\subsection{Design Iterations}
Our design team involved a senior UI designer familiar with search interfaces and literature support tools who provided feedback on the usability and clarity of our system design through the iterations. We also ran three rounds of pilots to seek design feedback and iterate before running the evaluation study. The iterations sought to improve clarity around the main confusion points discovered from pilots (described below) and usability (\eg adding a central state and in-line action indicators for loading latencies). See Appendix~\ref{appendix:design-iterations} for description descriptions \new{and design rationales}.


\subsection{Final System Interface}
The final interface is shown in Fig.~\ref{fig:interface}. \cirnum{A} \& \cirnum{B}: Saved authors and papers are shown in the corresponding panels on the left-hand side, to increase user's context awareness and allow them to track their progress over time. \cirnum{C} In each author recommendation header, pertinent information about the author such as their name\footnote{We collapse the first name to reduce subconscious focus on presumed gender of the author. However, users are instructed to mouse-over to see the full name of the author or click on the author's name to see additional details of that author on the corresponding author details page on Semantic Scholar.},
the number of their papers the user has saved or downvoted, the number of total publications and estimated relevant ones, their h-index, and the number of citations. Next to the author's name is a small histogram visualization of the author's publication records over time, with defeault blue overlay bars in the vis \new{showing} the \new{number} of papers predicted relevant by the system at the time of recommendation over the years. This visualization updates when the user clicks on a relevance explanation `pill' available under the author name (\cirnum{F}), by adding the corresponding counts of papers for the filter as an additional overlay \new{using} the same color. Clicking on a filter also filters the corresponding papers in the author's publications list (\cirnum{E}), sorted in a descending order of the predicted relevance score \new{(default)}. \cirnum{D} A stack of papers the user has already provided feedback on appears as a stack at the top. The user can provide feedback on each paper to save or downvote it, which updates the backend while adding the title of the corresponding paper to the saved papers panel (\cirnum{G}). In each paper recommendation, users can click on any author name to open an author details \new{modal (not shown)} that looks exactly like the \new{author recommendation cards} in the main tab (\cirnum{H}). \new{Users can} view who among the saved authors cited \new{each author's} papers (\cirnum{I}). In particular, this information `bubbles up' to the recommendation header (if the \new{citing} author is not featured in author-level `cited by' filters). By default \new{only 5} papers for an author \new{are shown} to prevent overload, \new{but is expandable} (\cirnum{J}). Finally, users can click on the `Load More Authors' button at the bottom to receive a new batch of recommendations (\cirnum{K}). Saved author names are highlighted in green in a paper recommendation context (\ie \new{they serve as \textit{explanation}} author highlights in contrast to \textit{recommendation} highlights \new{in the baseline}).


\subsection{Baseline} \label{section:baseline_system}
\begin{figure}[h]
    \vspace{-1em}
    \begin{subfigure}[t]{0.5\textwidth}
        \centering
        \includegraphics[height=2.5cm]{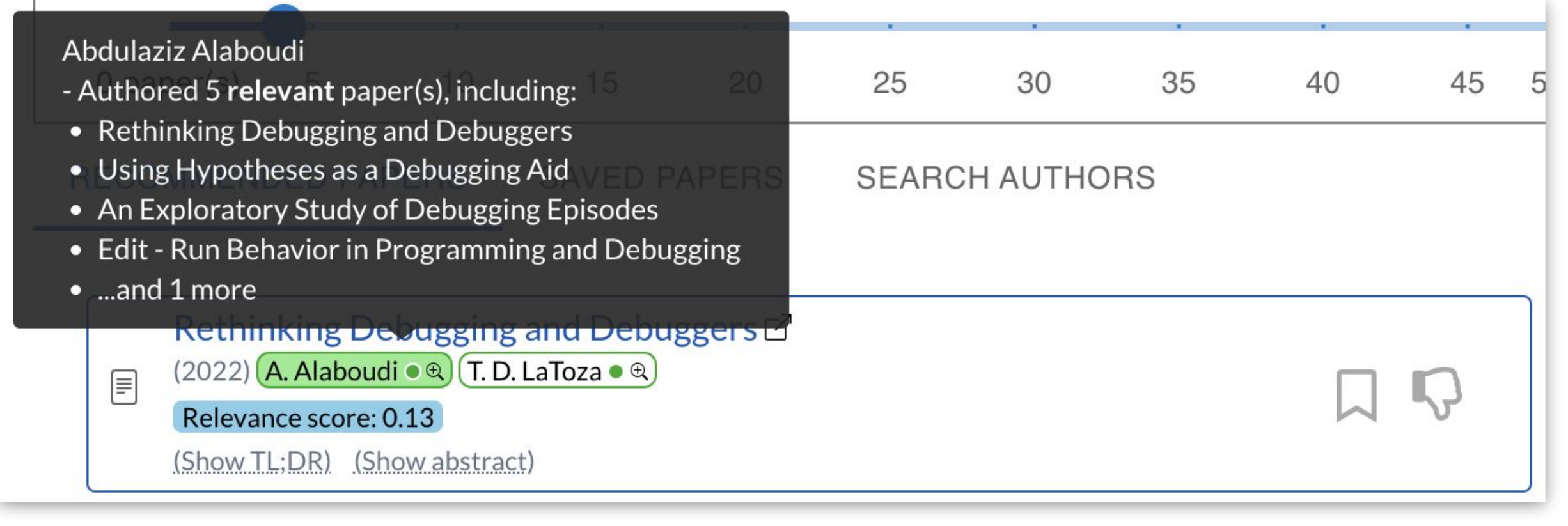}
        \caption{\new{Mousing} over an author name \new{reveals} a list of paper titles\\published by the author, predicted as relevant by the system.}
        \label{fig:baseline_tooltip}
        \Description{An image of a tooltip displaying a list of papers published by the author that the user mouse-over'ed that were also predicted as relevant by the system.}
    \end{subfigure}
    \quad
    \begin{subfigure}[t]{.45\textwidth}
        \centering
        \includegraphics[height=2.5cm]{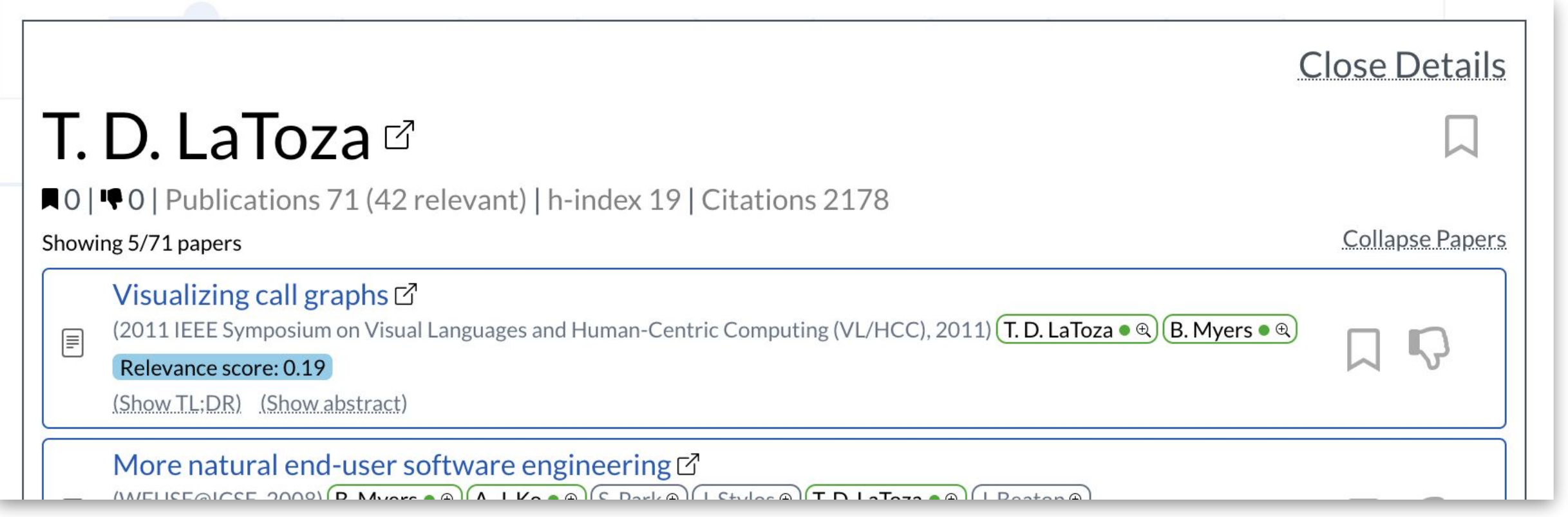}
        \caption{The author details page modal \new{is} accessible by clicking on an author name in any paper recommendation.}
        \vspace{-1em}
        \label{fig:author_details_page}
        \Description{An image shows the author details page modal which is accessible by clicking on an author name in any paper recommendation.}
    \end{subfigure}
    \caption{\new{Salient Baseline interface features.}}
    \vspace{-1em}
\end{figure}
\noindent The baseline and \sys interfaces \new{differed in the organization of recommendations}. \new{The top-level recommendations in the baseline featured a list of paper recommendations, with} an additional mouseover interaction for author names \new{in each paper} to reveal predicted relevant papers published by that author in a tooltip \new{when clicked} (Fig.~\ref{fig:baseline_tooltip}). Authors with a high number of relevant papers (\new{the threshold for highlighting was adjustable via} a slider at the top) featured a green dot (FeedLens~\cite{kaur-uist22-feedlens}) next to their names \new{with a highlighted border}. The baseline system instantiated a FeedLens mechanism in which it m\new{e}morized the paper lists the user encountered over time (\eg an author's publications when her detail's page is opened or the author is directly searched \new{by name}; when new paper recommendations are added to the main tab), while also scoring each paper in a list \textbf{using the same relevance prediction model implemented for \system}\new{, to update the author highlights in real-time}. Both interfaces supported click interaction \new{on author names} for opening a modal view of author details, including a \new{ranked} list of \new{author} publications \new{using} predicted relevance scores. In \sys author details pages featured more relevance explanation features (Fig.~\ref{fig:author_details_page}).

\vspace*{-4pt}
\section{Experimental Design}
\label{sec:evaluation-design}
\subsection{Objective \new{\& Research Questions}} Our goal in the evaluation was to study how \sys \new{and the author-centric interactions it instantiates} \new{benefit} scholars wanting to discover relevant and interesting {authors} and {papers} in a personalized domain. \new{Our research questions focused in part on the {efficiency} and {quality} aspects of scholars' literature discovery experience, for two modalities of discovery (\ie authors and papers). We operationalized the \textit{discovery efficiency} construct as the aggregated quantity of saved authors (or papers) for a fixed amount of time, \textit{quality} as the average post-task ratings of either relevance or interestingness of discovered items, and \textit{average discovered author novelty} as the ratio between the number of unfamiliar-yet-relevant authors to known-and-relevant authors. Concretely our research questions were:}
\begin{itemize}
\item \new{RQ1) Does \sys improve the efficiency and quality of scholars' author discovery over the baseline?}
\item \new{RQ2) Can \sys users save known-and-relevant authors and discover unfamiliar-yet-relevant authors?}
\item \new{RQ3) Comparing to a paper-centric baseline, does \sys inhibit paper discovery? and}
\item \new{RQ4) How do \sys users engage in paper discovery, and specifically does effort of discovery increase?} 
\end{itemize}

\vspace*{-4pt}
\subsection{\new{Participants}} We recruited 16 participants (8 female) for the study. The mean age of participants was 28.3 (SD: 4.32) and all actively conducted research at the time 
of the study (1 Post-doc, 13 PhD students, 2 Pre-doctoral Investigators). Participants' fields of studies included (multiple choices): HCI (9), NLP (6), Information Retrieval (2), Neuroscience (1), Oncology (1).

\vspace*{-4pt}
\subsection{Procedure}
\begin{figure*}[t]
    \centering
    \includegraphics[height=1.5cm]{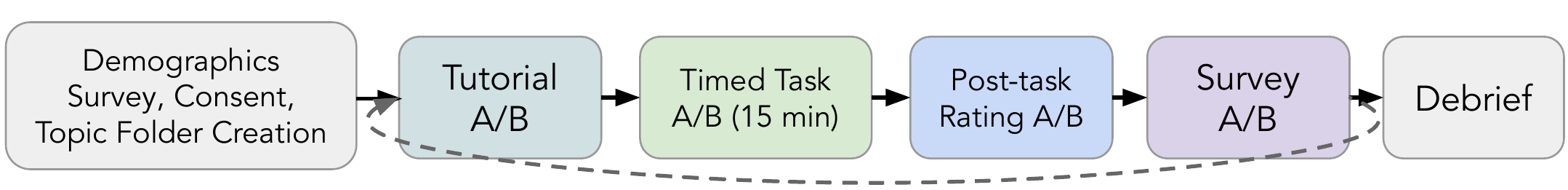}
    \vspace{-1em}
    \caption{The entire procedure of our study. The order of the middle section of the procedure was swapped based on the assignment (A/B). This order assignment was randomized and counterbalanced across participants (see text).}
    \Description{The process of evaluation.}
    \label{fig:study_procedure}
    \vspace{-1em}
\end{figure*}
\subsubsection{\new{Structure}} We employed a within-subjects study to compare \sys to a baseline system (see above for details of implementation). \new{We asked each participant to choose two different research topics they wished to explore, and randomly assigned systems to the topics for timed exploration tasks.} We counterbalanced the order of presentation using 8 Latin Square blocks and randomized rows. Participants followed the following procedure in the study, which took place remotely using Zoom (Fig.~\ref{fig:study_procedure}): Introduction, Consent, Demographics survey and curation of topic folders for the main tasks; Tutorial of the first system, Main task for the first system, Post-task rating, and Survey; Repeat for the second system; Debrief. In the topic folder curation, the interviewer guided each participant to navigate to a popular \new{online} scholarly search engine to create two topic folders, one per \new{participant's} research topic. At this stage, participant\new{s were} given time to freely search for two research papers that they thought represented each topic, and instructed to save them into the folders. \new{They} were asked to share their screen and think-aloud the main timed tasks. The study lasted \new{for} 1.5 hours and participants were compensated \$45 USD.
The study received Internal Review Board approval.

\subsubsection{Tutorials} 
Before participants start with each of the two main task with different conditions, they were given a tutorial of the assigned systems via screen sharing. The interviewer demonstrated the main features of each system based on a prepared script that took around 5 minutes for the baseline condition and 10 minutes for the system condition that had more features.
Participants were then instructed to save as many relevant and interesting authors and papers as possible during each task, and were recommended to save at least 5 in each category. Aside from it, they were also told to downvote 3 or more irrelevant papers early on to calibrate the recommender system.

\subsubsection{Timed Main Tasks (15 mins each)} The main tasks used the two different topics that participants chose as personally motivating for discovering new papers and authors in. We randomly assigned each topic to a condition. Each system used the two seed papers participants curated for each topic to generate the initial set of recommendations.

\subsubsection{Post-task Ratings and Surveys} \label{subsection:post_task_ratings}
After each task, participants clicked on a button in the interface to copy a random subsample of their saved authors and papers (up to 15, respectively, so that rating did not take overly long for any participant) and pasted this copied content onto a Google Spreadsheet that the interviewer shared with them. In the spreadsheet were three questions for each saved author and three questions for each saved paper. The first question for each author was a binary yes/no question (``\textit{Were you familiar with the author before the experience?}'') and the last two questions were 7-point Likert scale questions (``\textit{I found this author to be relevant.}'' and ``\textit{I found this author to be interesting.}''). For each paper, similar questions of familiarity, relevance, and interestingness were followed.

In the survey administered after each task, participants were asked about their subjective feelings related to the experience. For demand (including physical and cognitive) and overall performance we adopted the validated 6-item NASA-TLX scale~\cite{nasa_tlx}, with the original 21-point scale mapped to a compact 7-point scale~\cite{nasa_tlx_7point_use_case}. For technological compatibility with participants' existing discovery workflows and the ease of learning we adapted the Technology Acceptance Model survey from~\cite{tam_survey} (4 items). We also included additional questions to measure participants' subjective feelings of the system's effectiveness in supporting author (4 items) and paper discovery (3 items). Finally, we included additional questions for common (4 items) and condition-specific features (2 items in the baseline condition, 6 items in the treatment condition) of each system to measure their effectiveness (See Appendix~\ref{appendix:survey} for details of the questionnaire).

\subsubsection{\new{Data Collection}} During each participant's interaction with each system, we collected their behavioral traces \ie timestamped actions and \new{their} details. When \new{a} participant \new{provided} feedback on a paper, the unique paper identifier, its estimated relevance score from the recommender system at the time of feedback, and the context in which it appeared (\ie whether on an author detail's page) was stored. For each participant's feedback on an author, the unique author identifier, existing relations to saved authors at the time of feedback (treatment only), \new{and} in case of an author recommended in the main recommendations tab, which sourcing mechanism was used (treatment only), and whether the author was directly searched \new{was stored}. \new{We filtered the behavioral traces based on modality (\ie authors or papers) and transformed values (\eg average counts of saved authors; ratio between saved-to-downvoted papers) for analysis. Participants' think-alouds during the tasks, open-ended questions, and debrief were recorded and later transcribed.}

\subsubsection{\new{Data Analysis}} \new{The mappings between analyses of collected data and research questions are as follows.}
\begin{itemize}
\item \nnew{RQ1)} We analyzed the efficiency and quality measures of discovered authors \nnew{between conditions using the} paired Student's t-test. \nnew{We analyzed Likert items using non-parametric tests such as the paired-samples Wilcoxon's signed rank test (for paired-samples data such as participants' responses to survey questions) and the Mann-Whitney U test (for independent data such as judgment on saved authors or papers).}


\item \nnew{RQ2)} We analyzed the ratios between the number of discovered authors who were unfamiliar-yet-relevant (`unfamiliar') to the number of known-and-relevant (`known') authors. A mean novelty value $\in [-1, 1]$ was computed for each participant in each combination of experimental factors by averaging `unfamiliar' $\mapsto$ 1 and `known' $\mapsto$ $-1$ over saved authors, such that a value closer to 1 meant more unfamiliar authors were discovered for a unit number of known authors, and vice versa. We ran a one-way Repeated Measures (RM) ANOVA test with the experimental condition as a two-level factor (\ie \sys vs. baseline). RM ANOVA was chosen over regular ANOVA for its advantage in controlling for the random effect from subjects in the within-subjects experimental design. We tested the assumption of sphericity using Mauchly's test~\cite{mauchly1940significance} and ran post-hoc Tukey's HSD comparisons to identify significant pairwise differences. 

\item \nnew{RQ3)} We analyzed the efficiency and quality measures of discovered papers, similarly with RQ1.

\item \nnew{RQ4)} We analyzed both quantitative and qualitative data. The quantitative analyses included the average estimated model relevance scores of papers at the time of saving (the predicted relevance score on each paper ranged between $[-1, 1]$, where a positive score corresponded to relevance and vice versa, with higher significance towards both ends. This score represented how the recommender predicted the paper to be relevant, given all of the user's feedback on papers up to that point. The model for calculating the scores was held constant between the two conditions to control for analysis of the trends in user steering), the average number of papers saved for each discovered author, the balance of two steering operations performed on paper recommendations (\ie the mean ratio between the number of saving-to-downvoting was similarly calculated with the mean author novelty described above, by mapping `save' $\mapsto$ $1$; `downvote' $\mapsto$ $-1$, where a value closer to 1 represented a low net steering effort by a user for each saved paper and vice versa). We analyzed the average number of papers saved or downvoted for each author in each condition, using a two-way ANOVA (two two-level factors as experimental conditions and feedback type) followed by post-hoc Tukey's HSD tests; for analyses involving time progression, we ran RM ANOVAs as before. We checked the suitability of ANOVA by examining the homogeneity of variances in factor groups using Levene's test~\cite{levene1961robust}.
For qualitative analysis two authors analyzed transcripts through open coding, then discussed and merged main themes appeared from it.
\end{itemize} 


\section{Findings}
\subsection{\new{RQ1. }\sys increased author discovery efficiency \nnew{without decreasing} quality}
\subsubsection{Users saved more authors in \system and found saved familiar authors interesting}
\begin{figure*}[t]
    \includegraphics[width=.92\textwidth]{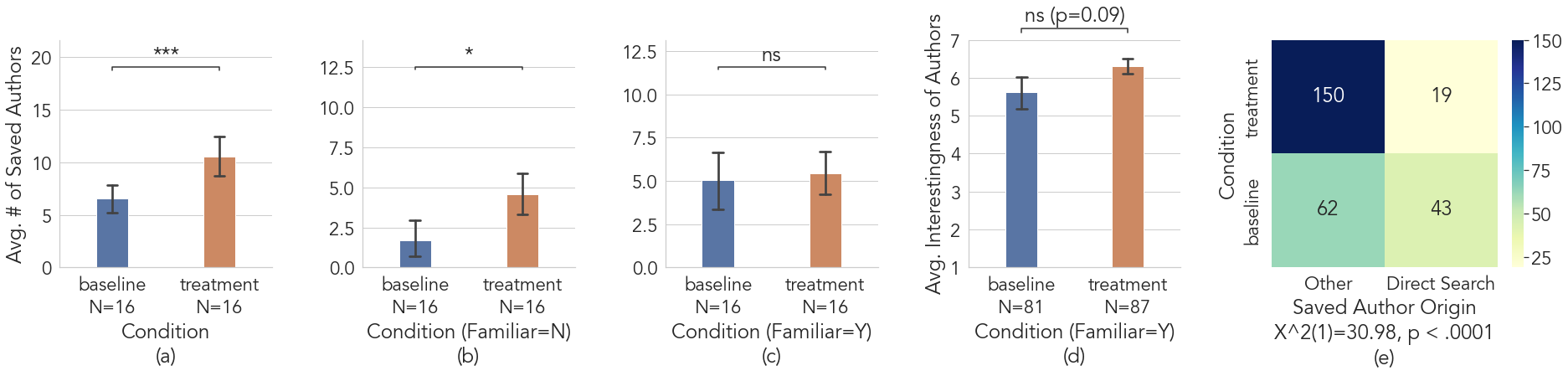}
    \vspace{-1.2em}
    \caption{Users' author discovery outcomes differed significantly between the conditions. (a) Users saved significantly more authors in \sys than the baseline \new{(the three tests in a--c were Bonferroni-corrected for multiple testing)}.
    (b \& c) The average number of unfamiliar authors saved was significantly higher in \system, whereas the average number of familiar authors saved did not differ between the two conditions. (d) The \nnew{difference of} average interestingness was \nnew{marginally significantly} higher for authors that users were familiar with prior to the task \new{but not for unfamiliar authors
    }. (e) Users in the baseline used direct search \new{with author names} significantly more, reflecting the primary means for finding familiar authors.}
    \vspace{-1em}
    \label{fig:author_saves}
    \Description{Four bar charts compare the baseline and treatment conditions in terms of the average number of authors users saved (a), divided by the saved author familiarity (b-c), and the subjective ratings of interestingness on familiar authors (d). A 2x2 matrix shows count distribution of saved authors' origins between the two conditions (e).}
\end{figure*}
Users saved significantly more authors overall in \sys (M=10.6, SD=3.88) than in the baseline condition \new{(at the $\alpha=.001$ level after correcting for multiple tests using the Bonferroni procedure,} M=6.6, SD=2.71, \tpaired{26.80}{-4.72}{\new{0.0003}}); Fig.~\ref{fig:author_saves}a). Between the familiar vs. \new{unfamiliar} authors who were saved, the distribution skewed towards familiar authors in the baseline condition, while a similar skew was not observed for the treatment condition (\chisq{10.86}{0.001}). This difference in distribution was reflected in the results of paired t-tests between the two conditions, with a significantly higher number of unfamiliar authors being saved in the treatment condition (M=4.6, SD=2.66; Baseline: M=1.7, SD=2.33, \tpaired{29.50}{-2.98}{\new{0.009}}, Fig.~\ref{fig:author_saves}b), while the number of familiar authors saved in each condition did not differ significantly (Treatment: M=5.4, SD=2.71; Baseline: M=5.1, SD=3.23, \tpaired{29.10}{0.33}{0.75}, Fig.~\ref{fig:author_saves}c), using a random subsample of authors \new{whose familiarity was} rated by users. Directed search (\new{\ie users typed in known author names}) was \new{significantly more common} in the baseline than the treatment condition, and this \new{was consistent with} how users saved significantly more unfamiliar authors \new{in \sys} where browsing and serendipitously discovering authors by clicking on their names in recommended papers was common (Fig.~\ref{fig:author_saves}e).

In terms of \new{the quality of discovery, measured as saved} authors' interestingness and relevance, both treatment (76\%=117/154) and baseline (84\%=79/94) conditions resulted in majority High interestingness (\nnew{See Fig.~\ref{fig:author_relevance_interestingness_hist} in Appendix~\ref{appendix:ratings-distributions}. for the aggregate response count distribution}). \new{In addition,} the overall distribution between High vs. Low interestingness authors did not differ significantly between the two conditions (\chisq{1.83}{0.176}). However, \nnew{we saw a marginally significant difference in interestingness between} the authors whom users were familiar with prior to the task in the treatment condition (M=6.3, SD=0.93) and the baseline condition (M=5.6, SD=1.97. \nnew{\mannw{3135}{0.089}},
Fig.~\ref{fig:author_saves}d), while no such difference was observed among the unfamiliar authors (\nnew{\mannw{1088}{0.80}}). We return to \nnew{this difference} in Section~\ref{subsection:qualitative_patches_of_research}\new{, RQ4}. The average relevance of \new{saved} authors did not differ significantly between the two conditions (\sys: M=6.4; baseline: M=6.4, \tind{161.93}{0.19}{0.85}).

\vspace*{-6pt}
\subsubsection{\new{Users found \sys features helpful for discovery}}
\label{section:survey_result} The survey results \new{corroborated these performance gains in \system}. The workload required to complete the task (measured via NASA-TLX) was significantly reduced in \sys (for \system, M=14.1, SD=5.56; for baseline M=17.0, SD=6.01, \nnew{\wilcoxon{13.5}{0.01}}). Users also responded that \sys better supported (a) author discovery: `helped me find relevant authors', M=6.1 (\system) vs. M=3.9 (baseline), \nnew{\wilcoxon{2.5}{0.002}}; `helped me make sense of author's research', M=4.8 (\system) vs. M=3.5 (baseline), \nnew{\wilcoxon{0.0}{0.008}}; `made me curious about author's research', M=6.1 (\system) vs. M=4.3 (baseline), \nnew{\wilcoxon{3.5}{0.001}}; `explanations of relevant authors became more helpful the more I used the system', M=4.9 (\system) vs. M=3.9 (baseline), \nnew{\wilcoxon{11.0}{0.02}} and (b) paper discovery: `helped me find relevant papers', M=6.1 (\system) vs. M=3.9 (baseline), \nnew{\wilcoxon{6.0}{0.002}}; `made me curious about the papers I found', M=6.2 (\system) vs. M=4.8 (baseline), \nnew{\wilcoxon{3.5}{0.01}}; `explanations of relevant papers became more helpful the more I used the system', M=4.6 (\system) vs. M=3.6 (baseline), \nnew{\wilcoxon{21.5}{0.05}} (see Table~\ref{table:full_survey2} for details). \new{Consistent with the perception of helpfulness, users favored} \sys in terms of the overall technology compatibility with their existing scholarly discovery workflows (for \system, M=22.6, SD=2.99; for baseline, M=19.4, SD=5.08, \nnew{\wilcoxon{24.0}{0.02}}) and the plausibility of future adoption (for \system, M=6.2, SD=0.75; for baseline M=4.9, SD=1.50, \nnew{\wilcoxon{4.0}{0.003}}, see Table~\ref{table:full_survey1} for details).

\vspace*{-6pt}

\subsection{\new{RQ2. \sys users saved familiar authors to scaffold subsequent discovery of unfamiliar authors}} \label{section:scaffold_subsequent_finding_of_unfamiliar_authors}
\begin{figure}[t]
    \centering
    \includegraphics[height=3.3cm]{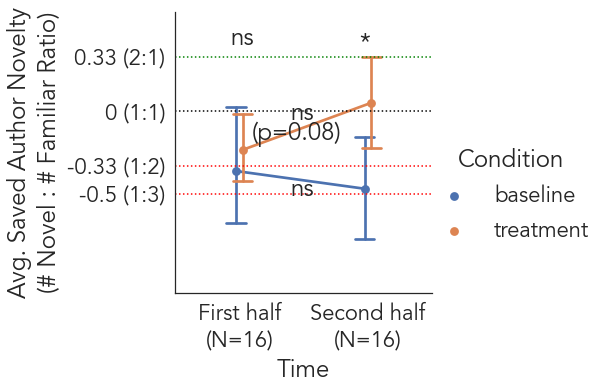}
    \vspace{-1em}
    \caption{\new{Though users in both conditions started with $\sim$$2\times$ more familiar authors to novel authors in the 1st half of the task,} users \new{in the treatment condition} saved \new{significantly more novel authors} in the \new{2nd half of the task, reaching familiar:novel parity}.}
    \vspace{-1.5em}
    \label{fig:author_save_familiarity_time}
    \Description{Users saved more novel authors in the second half of the experiment in the treatment condition than in the baseline condition.}
\end{figure}
Users' \textit{a priori} familiarity judgment on a random subsample of the saved authors showed that users in both conditions started with twice as many familiar authors as unfamiliar authors in the first half of the task, but in \sys users were finding more unfamiliar authors later on, nearing the parity between the number of familiar to unfamiliar authors in the second half (Fig.~\ref{fig:author_save_familiarity_time}; \nnew{As a measure of robustness, we examined possible variations in the number of saved authors, and found that they did not change significantly between the first and second half of the session; \chisq{0.03}{0.86}).}\new{The result of RM ANOVA showed a marginally significant overall effect $F(1,15)=4.23, p=0.057, \eta^2=.22$. Additional Tukey's pairwise HSD comparisons revealed that in the second half of the experiment users in \sys had a significantly higher ratio of novel authors saved than the baseline ($T=-2.66$, $p=0.01$, Cohen's $d=-.94$, partial $\eta^2=.18$). Time progression in \sys had a marginally significant ($T=-1.06$, $p=0.08$) positive effect (Cohen's $d=.64$, partial $\eta^2=.09$), but not in the baseline condition ($T=.17$, $p=0.87$). In terms of the sources of saved authors, all 3 mechanisms of recommendation seemed equally represented in the origins of these saved authors (a two-way ANOVA analysis showed no significant main effect from the recommendation mechanism type $F(2,90)=1.25, p=.29$, nor a significant interaction effect with the familiarity of saved authors $F(2,90)=1.29, p=.28$).}

\subsection{\new{RQ3. }\sys users saved more papers and rated saved papers as more interesting} \label{section:ratings_result}
\begin{figure*}[t]
    \includegraphics[width=.95\textwidth]{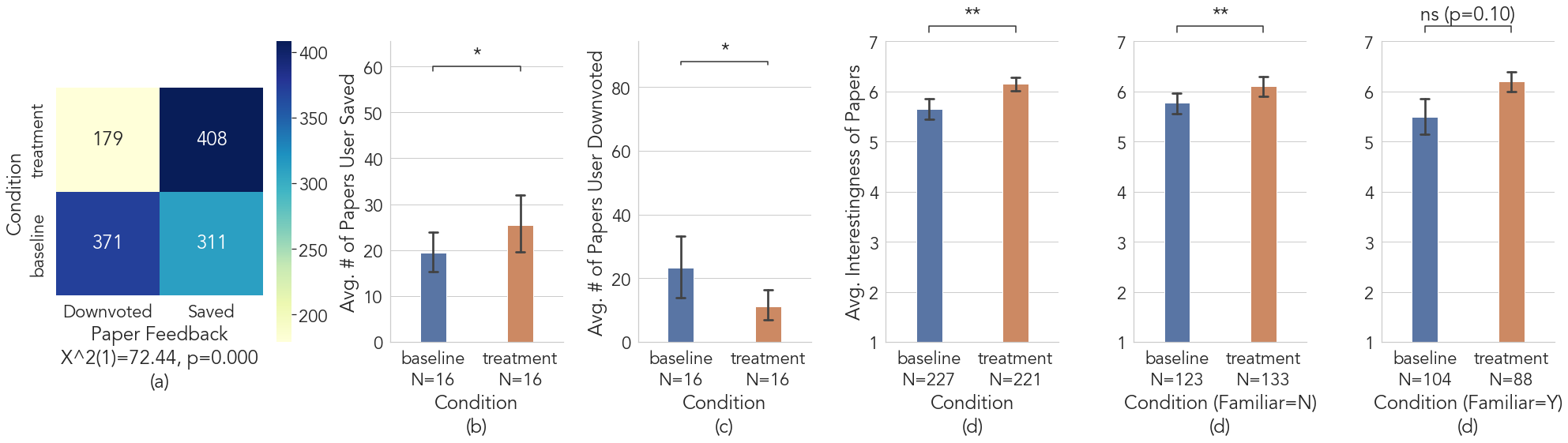}
    \vspace{-1.2em}
    \caption{Comparisons of user actions on and perceptions of paper recommendations between conditions. (a) The distribution of the number of saved vs. downvoted papers differed significantly, (b) with a skew towards saved papers in \sys (c) while more downvoted papers in the baseline. (d) The average post-task interestingness response on saved papers was significantly higher in \sys, (f) \new{but specifically} for papers that werenfamiliar\new{, while the average familiar paper interestingness did not differ significantly}. \new{Analyses in (d--f) were Bonferroni-corrected \nnew{for multiple testing (\ie three tests})}.}
    \vspace{-1em}
    \label{fig:papers_result}
    \Description{A 2x2 matrix shows the distribution of the number of positive versus negative feedback on papers between the two conditions (a). Five bar charts show the average number of papers users saved (b) or downvoted (c); the average interestingness judgment on papers (d); and divided into sub-groups by familiarity (e-f).}
\end{figure*}
\subsubsection{\new{Efficiency}} On average, \new{users} saved significantly more papers in \sys (M=25.5, SD=13.55) than in the baseline condition (M=19.4, SD=9.59, \tpaired{27.01}{-2.62}{0.02}; Fig.~\ref{fig:papers_result}b). Users in the baseline condition downvoted more papers (Treatment: M=11.2, SD=10.05; Baseline: M=23.2, SD=0.56, $t_{\text{paired}}(21.78)=-2.17, p=0.05$, Fig.~\ref{fig:papers_result}c)\new{, indicating an improved efficiency}. The distribution of \new{saved-to-downvoted papers} differed significantly, \chisq{72.44}{0.00}, Fig.~\ref{fig:papers_result}a). \new{Furthermore,} among the random sample of these papers rated by users post-task, participants in both conditions saved a similar number of familiar papers (Treatment: M=5.5, SD=3.43; Baseline: M=6.5, SD=3.97) and unfamiliar papers (Treatment: M=8.3, SD=3.40; Baseline: M=7.7, SD=3.70).
\subsubsection{\new{Quality}} The average (7-point Likert) interestingness response on saved papers was significantly higher in \sys (M=6.1, SD=1.07) than in the baseline condition (M=5.6, SD=1.54, $t_\text{two-tailed}$ $(403.00)$$=4.02$, Fig.~\ref{fig:papers_result}d). To \new{further investigate the distributional differences \nnew{in interestingness judgment}}, \new{w}e coded the response options 6 and 7 on the Likert scale as `High' interestingness, and the response options 4 and 5 as `Low' interestingness (\new{they} corresponded to \textit{moderate-to-strong} and \textit{neutral-to-slight} agreement \new{levels}, respectively. \nnew{See Fig.~\ref{fig:paper_relevance_interestingness_hist} in Appendix~\ref{appendix:ratings-distributions} for count distribution}). The resulting 2 (Interestingness) $\times$ 2 (Condition) matrix showed a significant skew towards High interestingness in both conditions, but with a higher degree in \sys (83\% of rated papers in \sys were judged as High vs. 73\% in baseline, \chisq{5.03}{0.025}). On average, familiar papers were judged \nnew{marginally} significantly more interesting in \sys (M=6.2, SD=0.91) than Baseline (M=5.5, SD=1.88, \nnew{\mannw{3954}{0.10}}, Fig.~\ref{fig:papers_result}f), \nnew{while for} unfamiliar papers \new{the difference was \nnew{significant}} (Treatment: M=6.1; Baseline: M=5.8, \nnew{\mannw{6622}{0.009}}, Fig.~\ref{fig:papers_result}e) \new{which users judged as similar relevance} (Treatment: M=5.7; Baseline: M=5.6, $p=0.41$).

\subsection{\new{RQ4. \sys helped `shortcutting' to more relevant papers, leading to efficiency gains and better human-AI alignment on relevance}}
\begin{figure*}[t]
    \begin{subfigure}[t]{.29\linewidth} 
        \centering
        \includegraphics[height=3.3cm]{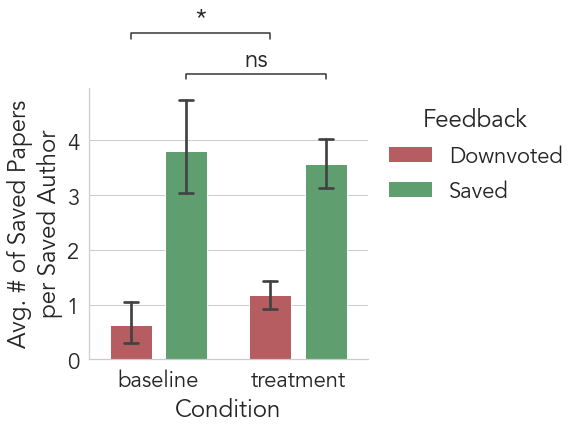}
        \caption{Users saved and downvoted multiple papers from each saved author's publications in both conditions, \new{suggesting how they recognized interrelated bodies of work.}}
        \label{fig:paper_save_from_author_context_baseline}
        \Description{Users saved and downvoted multiple papers from each saved author's publications in both conditions, suggesting how they recognized interrelated bodies of work.}
    \end{subfigure}
    \quad
    \begin{subfigure}[t]{.34\linewidth}
        \centering
        \includegraphics[height=3.3cm]{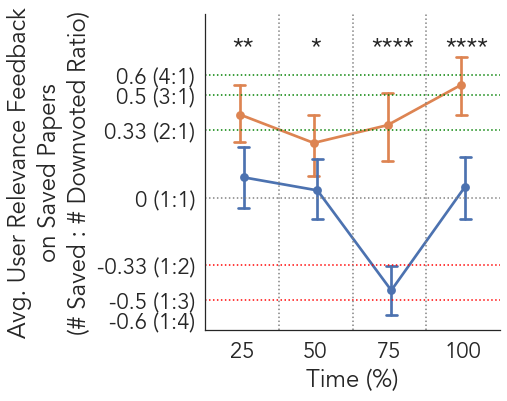}
        \caption{While in author-centric exploration, more than half of feedback was positive throughout the task, in paper-centric exploration users provided significantly more negative feedback in the third quarter of the task (see text).
        }
        \label{fig:paper_feedback}
        \Description{In the baseline condition users provided significantly more negative feedback in the third quarter of the task, representing an increased amount of user efforts in steering the recommender system, compared to the treatment condition which did not show such an increase in effort.}
    \end{subfigure}
    \quad
    \begin{subfigure}[t]{.31\linewidth}
        \centering
        \includegraphics[height=3.3cm]{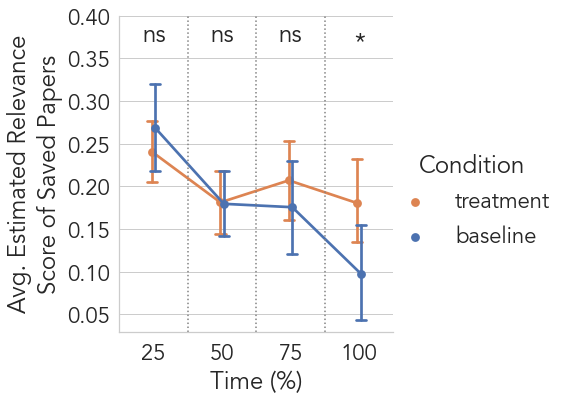}
        \caption{Predicted paper relevance scores \new{decreased on user-saved papers in the 1st half of the task}. However, \new{in the 2nd half} the average model relevance scores decreased significantly more in the baseline.}
        \label{fig:align}
        \Description{The user and the AI-based recommender system's predictions of relevance aligned better in the treatment condition than the baseline condition, as the experiment progressed, showing an increasing gap in the fourth quarter.}
    \end{subfigure}
    \vspace{-1.25em}
    \caption{(a \& b) Users actions and steering efforts; (c) Changes in machine-predicted relevance of saved papers.}
\end{figure*}
\subsubsection{\new{Users saved multiple papers from each discovered author at once}} Users \new{in both conditions} visited an author details page to find over 3 relevant papers at once and save them (Treatment: M=3.6, SD=2.96; Baseline M=3.8, SD=4.44, Fig.~\ref{fig:paper_save_from_author_context_baseline}). The number of downvoted papers was significantly lower than the number of saved papers in both conditions \new{(two-way ANOVA with Condition and Feedback Type as factors and the number of papers receiving feedback as a DV showed a significant main effect from Feedback Type: $F(1,542)=121.88, p<.0001$ but not from Condition ($p=.53$), nor from their interaction ($p=.12$)). The effect size of Feedback Type was Cohen's $d=-.98$, partial $\eta^2=.20$ (in \sys; $T=-9.05$, Tukey's $p=0.001$) and $d=-.93$, $\eta^2=.18$ (Baseline; $T=-6.69, p=0.001$)}. The result indicated the primary driver for navigating to a specific author's page to be finding relevant papers. However, the average number of papers downvoted in the treatment condition (M=1.2, SD=1.77) was significantly higher than in the baseline condition (M=0.6, SD=1.96, \new{Tukey's $p=0.02$}), suggesting that navigation to each author's publications was more motivated and contextualized in \sys to also recognize which threads of research were not relevant to the topic.

\subsubsection{\new{\sys users expended significantly less effort during paper discovery}} How much \new{effort users expended during their discovery} was evident in how much negative feedback they had to provide for the same number of saved papers\new{, in order to steer the system according to their changing notion of relevance}. Compared to \sys, baseline users provided much more negative feedback for a given number of saved papers (Fig.~\ref{fig:paper_feedback}). \new{The overall effect of experimental condition was significant, RM ANOVA $F(1,15)=21.51, p=.0003, \eta^2=.59$. Post-hoc pairwise Tukey HSD comparisons between the two conditions were significant in all four quarters, with the highest difference being the 3rd quarter: $T=-7.81$, $p=0.001$, Cohen's $d=-.89$, partial $\eta^2=.16$).}

\subsubsection{\new{Human-AI alignment on relevance was improved in \sys}} \new{The relevance model used for sourcing and sorting paper recommendations was held at constant between the two systems, hence examining its relevance scores (a score closer to 1 indicates higher relevance) calculated for \textit{saved} papers at the time of saving represents the degree of alignment between the human user's and AI's notions of relevance.} We found that system's predicted scores of relevance on papers saved by users exhibited a widening gap \new{on alignment} in the first half of the task as shown in the decreasing average estimated relevance scores for both conditions. However, in the second half of the task, the average predicted relevance scores of saved papers decreased more in the baseline condition, leading to a significant difference between the two conditions in the last quarter of the task (Treatment: M=0.18, SD=0.246; Baseline: M=0.10, SD=0.247, \new{post-hoc Tukey's HSD: $T=-2.14$, $p=.03$, Cohen's $d=-.33$, partial $\eta^2=.03$}, Fig.~\ref{fig:align}). 

\subsubsection{\sys users felt authors represented contextualized `patches' of relevant research} \label{subsection:qualitative_patches_of_research} \new{In a qualitative analysis we found themes that contextualize the results thus far.} Users felt that \sys helped them find unfamiliar-yet-relevant authors (\eg ``\textit{find a bunch of interesting authors that I didn't know about}'' -- P16
). In particular, P5 connected their experience to an analogy of foraging where ``\textit{[I would] drill down as looking at a specific author and the papers they publish... [it] helps me go from one world to another... like jumping from patch to patch.}'' Furthermore, \sys was perceived as ``\textit{providing more context}'' (P14) to exploration, helping users realize connections between two authors (``\textit{Now that's very interesting, because I didn't know these people were connected}'' -- P2) or discover earlier, or less familiar, work that they had not known for a familiar author: ``\textit{I know [Author] does other work that's not relevant to my interest but the signals (explanation features) really helped me tease out which of his work is relevant... and a lot of them I haven't read before. So I think this is great in terms of helping me discover some of his earlier work that I can find helpful.}'' (P10).
\section{Discussion and Future Work} \label{section:discussion_future_work}

\subsection{Study Design Limitations}
The study design makes several important trade-offs for practical considerations. First, post-task ratings is efficient to collect retrospective data about participants' experience during the study. However, they may also be subject to confirmation biases towards their own earlier judgements.
Second, conducting lab studies allowed us to control for unintended factors such as amount of time engaged with each systems. However, the relatively short duration of lab studies opens up the possibility that unobserved effects of time pressure may have \new{led participants to engage with} recommended items \new{in a shallow manner}. \new{Furthermore,} long\new{er} term effects of using the system requires a prolonged field deployment study \new{with} significantly more resource \new{demands}.
Both of these limitation should apply to both conditions equally.
Finally, to keep our study as realistic, we allowed participants to freely choose the two topics they wished to explore during the study to ensure the\new{ir engagement and prior knowledge}.
The trade-off here is that the two topics may have differed qualitatively along the dimensions of topical familiarity or the level of abstraction. To mitigate this, participants were instructed to think of topics at a similar level of abstraction (\eg headings in the related work section of their own papers), and the topics participants chose were randomly assigned to the conditions in the experiment. For these reasons we do not expect to see a significant confounding effect from differences between the two topics.

\subsection{\new{Technical and human factors design implications for future author-centric discovery systems}}
\begin{figure*}[t]
    \includegraphics[width=.95\textwidth]{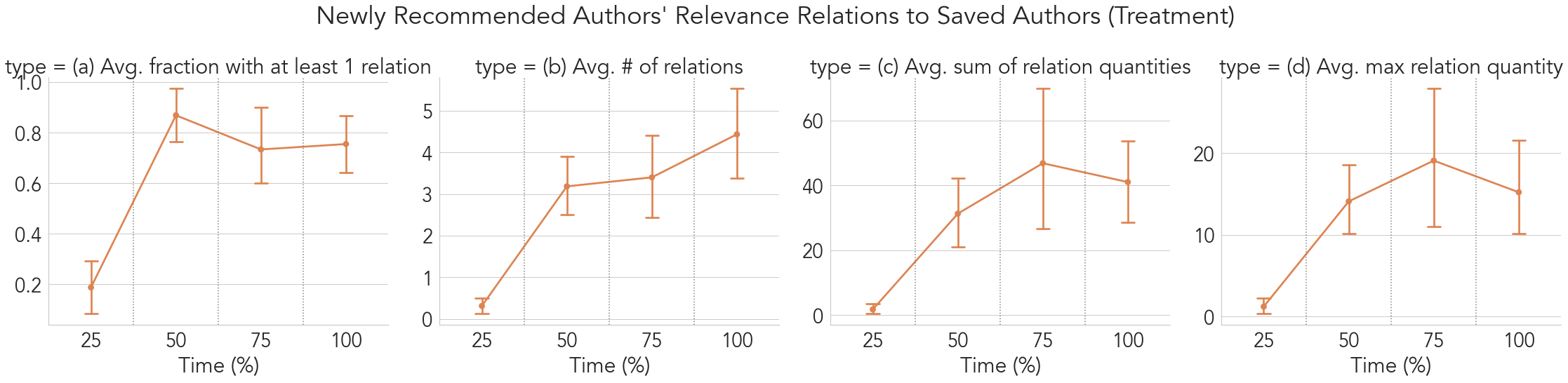}
    \vspace{-1em}
    \caption{(a, b) Initially author recommendations users saved did not \new{have} relations to saved authors (cold start), but \sys \new{could} quickly recommend authors with relations to existing saved authors, \new{to help} users rapidly form a \new{committee from the 2nd quarter and on} as shown in \new{the} significant increase in \% of authors with a relation.
    (c, d) \new{However, the utility of relation strength} (\ie the number of papers \new{that relate} a recommended author and a saved author) became marginal over time, suggesting that above a certain threshold participants did not need to differentiate the number of papers involved in a relation.}
    \label{fig:relevance_relation_over_time}
    \Description{Line charts show the existence of a cold-start phenomenon in the system but also the user's ability to rapidly combat it.}
\vspace*{12pt}
\end{figure*}
\subsubsection{\new{Latency of recommendations}} \label{section:latency} \new{The efficiency of discovery in \sys was observed in spite of its significantly longer latency for retrieving recommendations.} On average \new{a} recommendation request took significantly longer in \sys (M=12.9s, SD=8.69s) than Baseline (M=4.1s, SD=2.77s, \tind{89.58}{8.99}{\new{$3.73\times10^{-14}$}}), leading to a sizable difference in the numbers of user requests in each condition (N=83 in \sys vs. N=185 in baseline). Optimization could shorten this latency to be \new{conducive to} scal\new{ing} and longer term use (\eg \new{scoring and ranking authors' papers efficiently using pre-computed} summary embeddings~\cite{kaur-uist22-feedlens}). 

\subsubsection{\new{Combatting early cold-start phenomena}} Users in \sys attended \new{new authors'} relations to existing saved authors when saving \new{them}. As expected, initially author\new{s} users saved \new{featured few} relation\new{s} to saved authors \new{(}cold start\new{)}, as shown in the low \% of authors saved that had \textit{any} relation to a saved author (Fig.~\ref{fig:relevance_relation_over_time}a, M=19\%, SD=3.9\%) in the 1st quarter. However, in the 2nd quarter this quickly increased to 87\% (SD=3.4\%, \tind{82.24}{-8.56}{\new{$5.0\times10^{-13}$}}), and plateaued for the \new{remaining}. \new{T}he average number of relevance relations featured for \new{a} saved author, an indication of how strongly an author is related to \new{the} set of saved authors via coauthorship- and citation-based relations \new{at the time of saving}, also increased significantly from the 1st to 2nd quarter (from M=0.3, SD=0.72 to M=3.2, SD=2.28; \tind{42.87}{-7.49}{\new{$3.0\times10^{-9}$}}, Fig.~\ref{fig:relevance_relation_over_time}b). The quantity \new{of} each relation (\ie how many papers did the two authors coauthor?; \new{or} cite from \new{one an}other?) \new{for a saved author} increased from the 1st to 2nd quarter, and remained high (Fig.~\ref{fig:relevance_relation_over_time}c, d).
Taken together, \new{these results show that while cold-start may be a challenge for author-centric discovery systems, users can combat this by recognizing relevant authors through recommendations sourced via triadic closure on citation networks, augmented with interactive relevance explanations, and iteratively curating them.}

\subsubsection{\new{Getting stuck in a particular `school of thought' vs. steering the system to diversify discovery in later stages}}  \label{section:qualitative_analysis_challenges_of_steering_in_treatment} While users perceived \sys as helpful for making sense of scholarly relations among authors, on the flip side they paid more attention to and noticed more whether a new author recommendation belonged to a group of authors. Despite having saved more authors and especially unfamiliar \new{ones} compared to the baseline system \new{(RQ2)}, users felt they had \new{difficulty} steering \sys to recommend authors with more diverse research backgrounds. P5 reasoned this \new{as} ``\textit{Only a small group -- twenty to thirty -- authors that, like, really work in that space a lot, and... co-author a lot of stuff together so it's kind of easy to stay in that insular community of authors.}'' (P5). Users recognized potential dangers of ``\textit{falling into an echo chamber... where people that have the most papers are the ones given the most attention}'' (P15) and ``\textit{it could be kind of hard to get out of a close-knit group of authors because they are interconnected}'' (P7). P10 described \new{that}:
\begin{quote}
``\textit{[The system] is too good in recommending to a point where I had to fight with it a little in order to step out of the immediate circle... so I was a bit hesitant to add more authors from this school of thought because that will give the system even stronger signals... even if I wanted to step out of the immediate circle.}'' -- P10
\end{quote}
Users pointed out an interesting dimension of authors, seniority, and how it may be considered differently during the exploration: ``\textit{I liked seeing these (familiar) authors early on... but it makes you wonder, if I have a perfect search engine... would it recommend new or up-and-coming authors who published at less known venues later?} -- P5; ``\textit{So I blocked a bunch of (well recognized) authors towards the end who have broad interests, cited by everyone... compared to students who have more niche interests... and it (the system) was recommending less familiar names (after blocking them).}'' -- P11.

\new{Interestingly,} users also pointed out a \new{similar} steering challenge while interacting with the baseline, but with ``\textit{getting stuck in a bubble}'' (P16) at a paper, not author, level. This \new{in turn} may have negative downstream consequences on author discovery: ``\textit{Maybe I steered the model too much to what I already know... the papers I'm seeing are the ones that I've already read before... and therefore discovering new authors now is a little bit hard to do.}'' -- P16; ``\textit{It's interesting because I think going around in circles with the same group of authors or papers... it happened more in the (baseline system).}'' -- P12; ``\textit{It felt like the system kind of ran out of things to recommend.}'' -- P5. Taken together the challenges \new{around diversifying discovery with continued use, and especially for branching out of a particular school of thought} point to an interesting design implications for future systems aimed at supporting scholarly discovery of both authors and papers (Section~\ref{section:discussion_future_work}).

\subsection{Beyond triadic closure on citation networks} \label{section:discuss_triadic_closure}
Our work also contributes to the research in network analysis and social psychology on homophily~\cite{lazarsfeld1954friendship} and triadic closure (\cf~\cite{asikainen2020cumulative}).
Recent work by Abebe et al.~\cite{abebe_triadic_closure_network_integration} suggests that the forces of triadic closure on networks can have positive, desegregating potential to increase \new{overall} network integration. Here, we explored whether this insight also generalizes to scholarly discovery, and in doing so, \new{instrumented} an interactive system \new{to study user behaviors and} showed that they rapidly saved familiar authors early on, as a form of `navigational springboards' \new{(Section~\ref{section:scaffold_subsequent_finding_of_unfamiliar_authors})}, \new{to then} discover significantly more novel authors. \new{Despite this gain in novel author discovery,} users had concerns regarding branching out of a closely related group of authors in a later stage of using the system, \new{and in some cases} avoided saving more authors who were strongly connected to the \new{committee} (Section~\ref{section:qualitative_analysis_challenges_of_steering_in_treatment}). These findings demonstrate the feasibility of \new{leveraging} triadic closure on \new{authorship graphs} for \new{its} desegregating potential, by way of recommending authors connected via author\new{s familiar} to the user to then navigate them to save more unfamiliar authors. \new{Fruitful avenue for future work lies in understanding the criteria for effective system} designs \new{for surfacing} core navigational \new{author nodes}.


\subsection{Finding non-interacting bodies of literature}
A potential limitation of methods relying on citation networks is its reduced discoverability potential of work outside frequently co-cited bodies of literature, leading to a form of filter bubbles~\cite{filter_bubbles_www_recommender}. Yet, bursting filter bubbles can have outsized potential for catalyzing significant innovations~\cite{rzhetsky2015choosing} especially for domains less likely to interact with each other~\cite{swanson_undiscovered,chu2021slowed}. Recent work on analogical scientific inspirations~\cite{kang_augmenting_tochi} and product innovation~\cite{hope_scaling} show early evidence on how analogical relations between papers may be computationally extracted, thereby filling in a `discovery hole' from relying only on conventional approaches or citation-based mechanisms to scholarly search. In addition, alternative complementary approaches may use differences between knowledge domains more explicitly to induce cross-domain retrieval in recommendations (\eg~\cite{naacl2022_kang_augmenting}). This line of prior work therefore points to fruitful avenues for future extensions. 
\subsection{Diverse intentions of user feedback} Our findings suggest that potential misalignment between the user's and the system's model of relevance \new{need to be carefully handled.}
Our quantitative (Fig.~\ref{fig:align}), behavioral (Fig.~\ref{fig:paper_save_from_author_context_baseline},\ref{fig:paper_feedback}), and qualitative analyses suggest that this may be an important issue for user adoption and the utility of the system. 
In particular, users perceived that the system was converging to a specific type of similarity and hypothesized that this was due to its optimization objective in both conditions, which was described as ``\textit{going around in circles with a close-knit group of authors}'' in \sys and ``\textit{papers topically too similar in nature}'' in the baseline system.
\new{Potential} implications for future systems that aim to support the changing notion of user relevance and their alternating desires -- sometimes seeking broader and other times more focused results -- are around how various user intentions of feedback may be supported, such as 1) steering the recommendations with varying degrees of expected changes in the outcomes (\ie small amount of feedback for tuning vs. large amount for jumping); 2) saving interesting (but not necessarily relevant to the task) intermediate results; and 3) subselecting from the user's own feedback to \new{experiment with} model behaviors, form a mental model, and re-use the subselection as a query or a sub-folder of the topic in case good results are yielded. 

\subsection{Flow of time} Our findings also uncover how time is an important dimension of literature discovery, and \new{especially} for systems that aim to leverage users' prior knowledge early in the interaction. The users of \sys exhibited a distinctive behavioral pattern that consisted of rapidly leveraging known, familiar authors early on (Fig.~\ref{fig:relevance_relation_over_time}), and seeking novel authors afterwards (Fig.~\ref{fig:author_save_familiarity_time}). Successfully adapting the system's objective in relation to this temporal context could improve user perception of its alignment and subsequently adoption. Some users explicitly commented on how they later blocked several of the more senior and well-known authors they found relevant earlier, in hopes of targeting authors with ``\textit{more specified, niche research interests.}'' Therefore future systems may design for the ability to filter outcomes based on certain attributes in a post-hoc manner, or prioritize items that meet certain criteria at the recommendation time. Other users also hypothesized that accumulation of their feedback over time made effecting desired changes in outcomes require commensurable amounts of feedback, which felt laborious and potentially frustrating in repeated use. They expressed the need to ``\textit{tell the machine to forget about my early feedback, without having to go back and redo it myself.}'' While a long line of research exists for modeling user's cognitive state, here we emphasize how simple interaction affordances such as being able to pull up the history of user feedback, locate a time range of their feedback, and specify the type\new{s} of feedback no longer relevant may prove to be effective. Finally, yet others commented on how they thought they were saving authors who are disproportionately more well-known or familiar using \system, even though the results showed that users overall saved more, not less, unfamiliar authors. This implies users may want to actively reflect on their progress over time, and making it explicit may help to \new{minimize} the potential perception biases \new{that may increase due to} sensitivity to certain attributes of the saved results.
\subsection{Supporting application designs using artifacts of exploration} While our evaluation showed the feasibility of integrating author-centric organization with paper recommendations to support users in rapidly building mental models of the literature, this also opens up many possibilities for future application designs that build on top of the artifact users curated during the exploration. One such example is interactive, \new{affinity-based} grouping of saved authors. Our users alluded to this possibility by commenting on how ``\textit{different schools of thought}'' co-exist in the literature which was made visible from looking at explanations of authors' citation and coauthorship relations, and how they might want to break out of one school to another for their own exploration. Incorporating additional attributes of authors may help with identifying different affinity signals related to this. For example users described how disciplinary background of an author may represent a specific kind of frequently used epistemological approaches, and how being able to group them based on this axis would contribute to understanding how different approaches \new{in a topic} evolved over time and interact with each other.
Another example application design is to extend the publication time visualization featured in \sys to groups of authors as a means to communicating which areas of research and epistemological approaches have recently been popular, and thereby empowering users to see what may be yet-to-be discovered or overlooked areas of future research opportunities. Our users also alluded to this possibility by commenting on how certain literature is `sparser' or `more densely published' than others, and how interesting spaces for interdisciplinary approaches can be found from looking at the adjacent areas of literature that they may pull from\nnew{, which suggests a fruitful avenue for future application designs that employ interactive visualization techniques to support effective user exploration.}
\section{Conclusion}
An important aspect in the domain of supporting scholarly discovery via interactive systems, which has not received significant attention from researchers, is how scientists' evolving knowledge of others may be captured and utilized to enhance their experience of discovering new authors and papers. Here we reflect on the design space of such tools and introduce \system, a literature discovery system that supports enhanced author-centric exploration by first enabling user curation of relevant authors, second using the curated authors to compute relevance signals on other authors and their work, and lastly using these signals to recommend further relevant authors and enhance understanding of the recommendations. We demonstrate the feasibility and value of \sys in a controlled study, and specifically show how \sys leads to discovering a larger number of relevant papers and authors \nnew{in a given amount of time, and how} participants also rated \nnew{discovered authors as more novel, while discovered papers} as more interesting. \nnew{Furthermore, we show} how initially forming a group of authors familiar to the user can lay down an enriched path for subsequent discovery.

\begin{acks}
This project is supported by NSF Grant OIA-2033558. The authors thank Luca Soldaini and Chris Wilhelm for their advice on and help with engineering the system; Daniel S. Weld, Doug Downey, and the researchers in the Semantic Scholar team for discussions and thoughtful feedback on the project. We also thank the anonymous reviewers for their constructive feedback. Finally, this work would not have been possible without our pilot test and user study participants. %
\end{acks}

\bibliographystyle{ACM-Reference-Format}
\bibliography{main}

\appendix
\section{How author recommendations were sorted in a batch for presentation} \label{appendix:sort-authors}
The position of an item on a list may determine whether it receives user's meaningful attention. While the batchified exploration in our system did not serve a large quantity of recommendations at once (up to 8 items), it nonetheless needed an order of presentation \new{among the items in each batch}. We iterated on its design through pilot \new{studies}. Our initial ordering simply interleaved the different sourcing mechanisms, which led to stratified sampling. Participants commented that this was confusing due in part to how author recommendations with several relevance explanation filters were featured lower than those that had none, \new{because they expected a higher relational strength} (for example, this happened when the author recommendations from the citation-based expansion mechanism were assigned to appear later than those from the relevant paper recommendations-based mechanism). To prevent this, we defined a relevance ratio as follows. For each author $a_i$, and her author-level relevance tags $\tau_j(a_i)$, we count the number of unique papers $C_p$ that appeared in them: $\sum_j C_p(\tau_j(a_i))$. To further enrich the strength of the relevance signal while also making comparison between authors fairer (\ie authors who have \textit{less} total publications but \textit{more} related publications out of the total may be perceived to have a higher density of relevant work, and therefore more interesting to the user), we normalized this quantity by the number of publications: $\sum_j C_p(\tau_j(a_i))/C_p(a_i)$.

\section{Pseudocode of author recommendation} \label{appendix:algorithms}
Pseudocode for author recommendation is shown in Algorithm~\ref{pseudocode:algorithms}.
\begin{algorithm*}
\caption{\new{Pseudocode Descriptions of Co-authorship- and Citation-based Recommendation Algorithms}}
\label{pseudocode:algorithms}
\begin{algorithmic}[1]
\Procedure{Sort-Sample}{$S$, $k$, $N$} \Comment{Approximating the most topically relevant papers for efficiency}
    \State Sort $s \in S$ in a descending order of \Call{Accessor}{$s$, $k$} \Comment{\Call{Accessor}{$s$, $k$} returns value of the field $k$ from $s$}
    \State Sample $p_1,\cdots,p_N$ from the top of the sorted list
    \State \Return $\{p_1,\cdots,p_N\}$
\EndProcedure
\Procedure{Vote-Multi}{$P$} \Comment{Each paper adds 1 vote to each of its authors}
    \State $\Omega \gets \text{Empty Dictionary}\ \omega_{\varnothing}$ \Comment{$\Omega:=$ (key: Author ID, value: Frequency $\geq 0$) store}
    \For{$\forall p \in P$}
        \For{$\forall a \in A_p$} \Comment{$A_p:=$ Authors of \nnew{Paper} $p$}
            \State $\Omega[a_{\text{ID}}]$ += 1 \Comment{$a_{\text{ID}}:=$ ID of author $a$}
        \EndFor
    \EndFor
    \State \Return $\Omega$
\EndProcedure
\Procedure{Get-Relevant-Papers}{$A, P_{\text{feedback}}$}
    \State $P_{\text{filtered}} \gets \varnothing$
    \For{$\forall a \in A$}
        \For{$\forall p \in P_a$} \Comment{$P_a:=$ Publications of \nnew{Author} a}
            \If{$p \notin P_{\text{feedback}}$}
                \State $p_{\text{score}} \gets$ \Call{Score}{$p, P_{\text{feedback}}$} \Comment{\Call{Score}{$p, P_{\text{feedback}}$} returns score of $p$ with ensemble SVMs}
            \Else
                \State $p_{\text{score}} \gets$ \Call{Retrieve}{$p, P_{\text{feedback}}$} \Comment{\Call{Retrieve}{$p, P_{\text{feedback}}$} returns user-feedback on $p \in \{-1, 1\}$}
            \EndIf
            \If{$p_{\text{score}} > 0$}
                \State $P_{\text{filtered}} \gets P_{\text{filtered}} \cup \{p\}$
            \EndIf
        \EndFor
    \EndFor
    \State $P_{\text{sampled}} \gets$ \Call{Sort-Sample}{$P_{\text{filtered}}$, score, 100}
    \State \Return $P_{\text{sampled}}$
\EndProcedure
\Procedure{Co-authorship-based Recommendation}{$A$, $P_{\text{feedback}}$} \Comment{$A:=$ Committee authors}
    \State $P_{\text{sampled}} \gets$ \Call{Get-Relevant-Papers}{$A, P_{\text{feedback}}$} \Comment{$P_{\text{feedback}}:=$ Papers with user feedback}
    \State $\Omega \gets$ \Call{Vote-Multi}{$P_{\text{sampled}}$}
    \State $\Omega \gets \Omega \setminus \{\text{self}, \forall a \in A\}$ \Comment{\nnew{Remove the user herself and committee authors}}
    \State \textbf{return} $\Omega$
\EndProcedure
\Procedure{Vote-Author}{$P$, $A$} \Comment{Each paper receives up to 1 vote (when cited) from each author}
    \State $\Omega \gets \text{Empty Dictionary}\ \omega_{\varnothing}$ \Comment{$\Omega:=$ (key: Paper ID, value: Frequency $\geq 0$) store}
    \For{$\forall a \in A$}
        \For{$\forall p \in P$}
            \If{$a$ cites $p$}
                \State $\Omega[p_{\text{ID}}]$ += 1 \Comment{$p_{\text{ID}}:=$ ID of paper $p$}
            \EndIf
        \EndFor
    \EndFor
    \State \Return \Call{Sort-Sample}{$\Omega$, ID, 100}
\EndProcedure
\Procedure{Citation-based Recommendation}{$A$, $P_{\text{feedback}}$} \Comment{$A:=$ Committee authors}
    \State $P_{\text{sampled}} \gets$ \Call{Get-Relevant-Papers}{$A, P_{\text{feedback}}$} \Comment{$P_{\text{feedback}}:=$ Papers with user feedback}
    \State $P_{\text{cited}} \gets$ \Call{Get-References}{$P_{\text{sampled}}$} \Comment{{Returns a set of referenced papers from $\forall p \in P_\text{sampled}$}}
    \State $P_{\text{voted}} \gets$ \Call{Vote-Author}{$P_{\text{cited}}, A$}
    \State $\Omega \gets$ \Call{Vote-Multi}{$P_{\text{voted}}$}
    \State $\Omega \gets \Omega \setminus \{\text{self}, \forall a \in A\}$ \Comment{\nnew{Remove the user herself and committee authors}}
    \State \textbf{return} $\Omega$
\EndProcedure
\end{algorithmic}
\end{algorithm*}


\section{Users' actions represented authentic relevance and interestingness} \label{appendix:authentic-user-actions}
\begin{figure*}[t!]
    \centering
    \includegraphics[height=3.1cm]{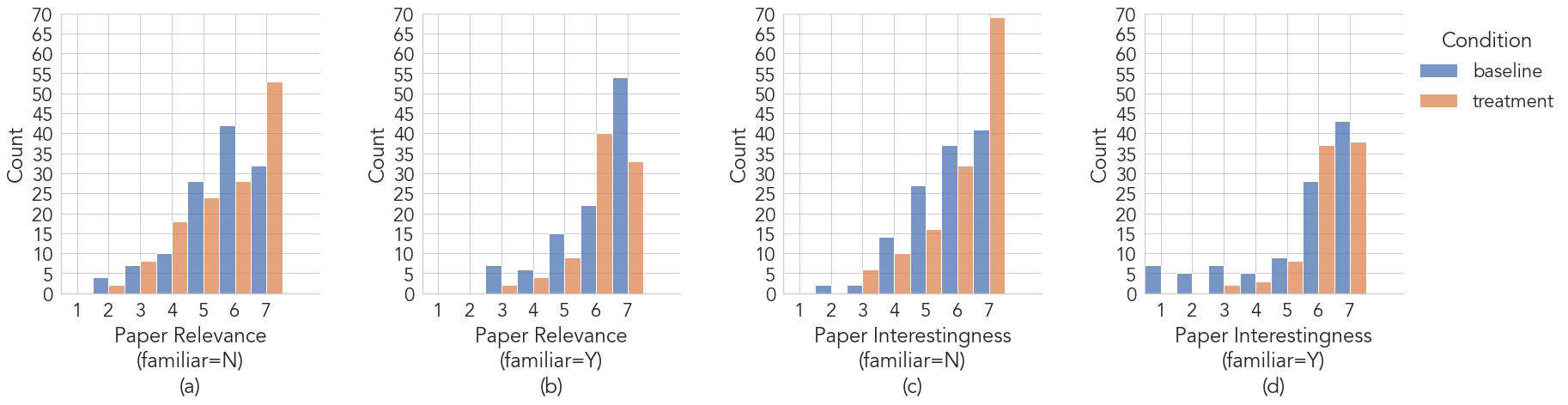}
    \vspace{-1em}
    \caption{Histogram of (a \& b) post-task paper relevance ratings for either familiarity type and (c \& d) paper interestingness.}
    \vspace{-1em}
    \label{fig:paper_relevance_interestingness_hist}
    \Description{Histogram of (a \& b) post-task paper relevance ratings for either familiarity type and (c \& d) paper interestingness.}
\end{figure*}
\begin{figure*}[h]
    \centering
    \includegraphics[height=3.1cm]{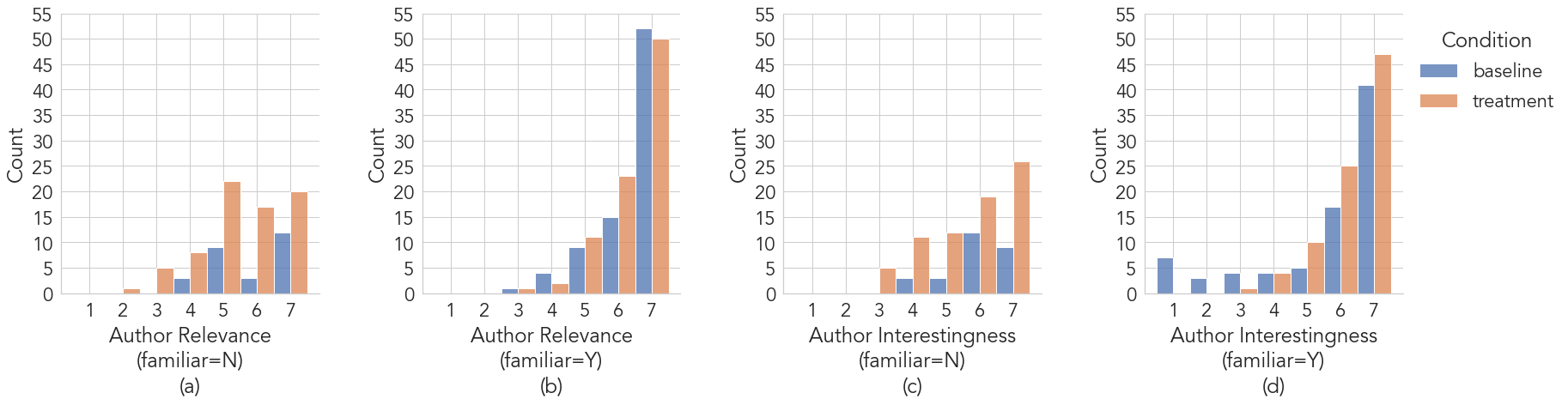}
    \vspace{-1em}
    \caption{Histogram of (a \& b) post-task author relevance ratings for either familiarity type and (c \& d) author interestingness.}
    \vspace{-1em}
    \label{fig:author_relevance_interestingness_hist}
    \Description{Histogram of (a \& b) post-task author relevance ratings for either familiarity type and (c \& d) author interestingness.}
\end{figure*}

\begin{figure*}[t!]
    \begin{subfigure}[t]{.45\textwidth}
        \centering
        \includegraphics[height=1.05cm]{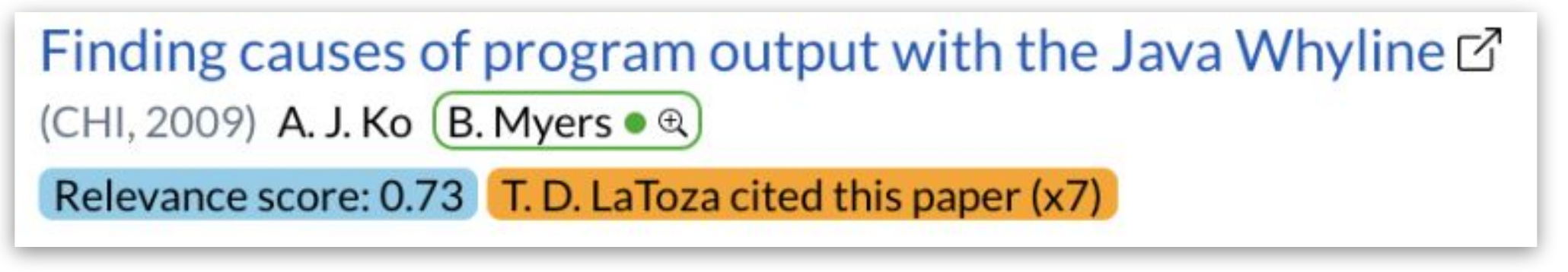}
        \vspace{-.5em}
        \caption{An example paper recommendation featuring a ``cited by [a saved author]'' label. Self-citations were excluded (see text).}
        \label{fig:removing_self_cites}
        \Description{An example paper recommendation featuring a ``cited by [a saved author]'' label. Self-citations were excluded.}
    \end{subfigure}
    \quad
    \begin{subfigure}[t]{.45\textwidth}
        \centering
        \includegraphics[height=1.05cm]{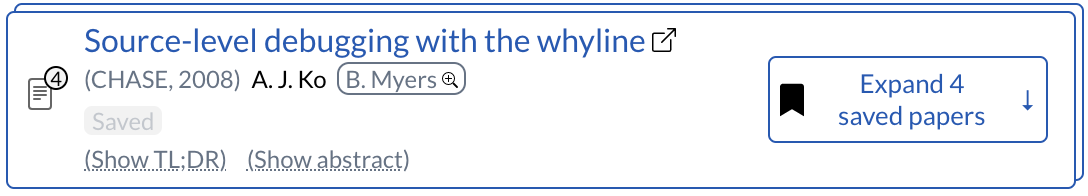}
        \vspace{-.5em}
        \caption{An example `stack of (judged) papers' UI for balancing the need for seeing familiar vs. new papers for an author (see text).}
        \label{fig:stack_of_papers}
        \Description{An example `stack of (judged) papers' UI for balancing the need for seeing familiar vs. new papers for an author.}
    \end{subfigure}
    \vspace{-1em}
    \caption{\new{UI designs in the baseline interface}} 
    \vspace{-1em}
\end{figure*}
Importantly, few of the randomly sampled saved papers were judged as irrelevant (on average users in both conditions rated around 13 out of 15 randomly sampled papers as 4 or higher on the relevance scale), suggesting that users' \new{save-paper} actions represented their authentic judgment of relevance. We also validated the significant difference in the average number of saved authors between the conditions by using user-vetted relevance ratings on saved authors. For \new{save-author} actions, \new{we ran an additional validity check} by removing authors rated as 4 or \new{lower} on the relevance scale\footnote{\Ie 4 corresponded to neutral agreement on `\textit{I found this author to be relevant}.' This presents a possibility that the author may become relevant had there been more time to explore. For this reason, we include response 4 as an indication of relevance. However we still observe a statistically significant difference between the conditions when authors scored 4 are excluded in the analysis (\tpaired{28.00}{-2.88}{0.01}).}, \new{and re-analyzing trends in the number of saved authors}. We see a consistent trend where users saved significantly more authors in \sys (M=9.6, SD=2.63) than Baseline (M=6.7, SD=2.75, \tpaired{29.94}{3.71}{0.002}), \new{suggesting} that users' decisions to save an author similarly represented their authentic judgment of relevance of the author. Taken together, we conclude that user save actions \new{likely} represented a level of authentic user interest and relevance in the saved items, beyond \new{merely as a means to steering} the recommender system.

\section{Distribution of post-task ratings on saved papers and authors, by familiarity}
\label{appendix:ratings-distributions}
\new{Distribution of the counts on paper and author ratings are shown in Fig.~\ref{fig:paper_relevance_interestingness_hist} and Fig.~\ref{fig:author_relevance_interestingness_hist}, respectively.}
\section{Description of design iterations} \label{appendix:design-iterations}
\xhdr{Removing self-citations from relevance explanations} Pilot users expressed  their intent for clicking on a relation explanation filter
\new{of an} author (\ie ``cited by [a saved author]'') \new{as} to see other papers by the saved author that cited the recommended author's papers. When these papers included self-citations, however, users did not feel it \new{matched their intent and question} the \new{usefulness} of relations due to \new{the self-promoting nature of self-citations}. To \new{align with the user intent}, we excluded self-citations from the data for featuring author-level explanations, and also from paper-level citation explanation labels for consistency (Fig.~\ref{fig:removing_self_cites}).

\xhdr{Increasing the information density on an author details page by adaptively minimizing judged papers} Pilot users also expressed wanting to see other papers by an author that they had not seen before first and foremost, rather than seeing papers that they had already provided feedback on. This makes sense especially in cases when the user is receiving new publication recommendations from authors whose earlier work they are already familiar with. However, in short-term use scenarios, we anticipated that there may be tension with users wanting to see familiar papers to build a mental model of and increase their confidence in judgment for a new author recommendation, especially when the user is trying to make a decision to save or downvote a paper. Therefore, we approached this trade-off by designing a mechanism for collapsing the familiar papers -- sorted from most to least recently interacted with -- that the user has provided feedback on into a stack of papers UI at the top of the author's publications list (Fig.~\ref{fig:interface}d), while also featuring an `expand' button next to the stack in case the user wanted to see individual papers in the stack (Fig.~\ref{fig:stack_of_papers}).

\xhdr{Presentation order of recommendations} Pilot users also \new{pointed out the} prominen\new{ce of the} `predicted number of relevant papers' tag featured for each author recommendation and \new{how it could be misleading when} author recommendations higher on the rank did not feature a higher \new{quantity}. Because this number \new{was perceived} useful for pilot users, but was not the main determinant of the \new{presentation} order of author recommendations \new{within} a batch (see Section~\ref{section:batch_generation}), we moved this information and made it less prominent in the final interface design (Fig.~\ref{fig:interface}c).

\section{Additional survey results}
\label{appendix:survey}
Descriptions of survey items and participants' responses grouped by condition are presented in Table~\ref{table:full_survey1} and \ref{table:full_survey2}. Two-sided paired samples t-tests were performed to compute the $p$-values between conditions. See Section~\ref{section:survey_result} for discussions of the results.

\begin{table*}[t!]
    \centering
    \begin{tabular}{p{2.25cm} p{6.2cm} p{2.0cm} p{2.0cm} p{.85cm}}
    \toprule
    & \textbf{Description} & \textsc{Baseline} & \sys & \textit{p}-val. \\
    \midrule
    \multirow{4}{*}{{{1. NASA-TLX}}} & Sum of the participants' responses to the five NASA-TLX's~\cite{nasa_tlx} Likert-scale questionnaire items below. The original 21-point scale was mapped to a 7-point scale, similarly with~\cite{nasa_tlx_7point_use_case}. & \multirow{4}{*}{17.0 (SD=6.01)} & \multirow{4}{*}{14.1 (SD=5.56)} & \multirow{4}{*}{\nnew{$.01^{*}$}} \\
    \midrule
    1a. Mental & ``How mentally demanding was the task?'' & 3.6 (SD=1.55) & 3.4 (SD=1.55) & \nnew{$.79$} \\
    \addlinespace[.1cm]
    1b. Physical & ``How physically demanding was the task?'' & 3.9 (SD=1.54) & 2.4 (SD=1.21) & \nnew{$.002^{**}$} \\
    \addlinespace[.1cm]
    1c. Temporal & ``How hurried or rushed was the pace of the task?'' & 3.1 (SD=1.69) & 2.7 (SD=1.40) & \nnew{$.38$} \\
    \addlinespace[.1cm]
    \multirow{2}{*}{1d. Effort} & ``How hard did you have to work to accomplish your level of performance?'' & \multirow{2}{*}{3.5 (SD=1.26)} & \multirow{2}{*}{3.1 (SD=1.54)} & \multirow{2}{*}{\nnew{$.33$}} \\
    \addlinespace[.1cm]
    \multirow{2}{*}{1e. Frustration} & ``How insecure, discouraged, irritated, stressed, and annoyed were you?'' & \multirow{2}{*}{3.0 (SD=1.59)} & \multirow{2}{*}{2.4 (SD=1.50)} & \multirow{2}{*}{\nnew{$.20$}} \\
    \addlinespace[.1cm]
    \hline\hline
    \addlinespace[.1cm]
    \multirow{5}{*}{{{2. TAM}}} & Sum of the participants' responses to the 4 questionnaire items below adopted from~\cite{tam_survey} measuring the technological compatibility with participants' existing scholarly discovery workflows and the easiness of learning. & \multirow{5}{*}{19.4 (SD=5.09)} & \multirow{5}{*}{22.6 (SD=2.99)} & \multirow{5}{*}{\nnew{$.02^{*}$}} \\
    \midrule
    \multirow{4}{*}{2a. Compatibility} & ``\textit{Using the system is compatible with most aspects of how I search for scholars and their papers.}'' (The response Likert scales for this question and below are 1: \textit{Strongly disagree}, 7: \textit{Strongly agree}) & \multirow{4}{*}{4.4 (SD=5.08)} & \multirow{4}{*}{5.1 (SD=1.24)} & \multirow{4}{*}{\nnew{$.15$}} \\
    \addlinespace[.1cm]
    \multirow{2}{*}{2b. Compatibility} & ``\textit{The system fits well with the way I like to search for scholars and their papers.}'' & \multirow{2}{*}{4.3 (SD=1.65)} & \multirow{2}{*}{5.1 (SD=1.02)} & \multirow{2}{*}{\nnew{$.14$}} \\
    \addlinespace[.1cm]
    2c. Easy-to-Learn & ``\textit{I think learning to use the system is easy.}'' & 5.8 (SD=1.05) & 6.2 (SD=1.02) & \nnew{$.12$} \\
    \addlinespace[.1cm]
    \multirow{2}{*}{2d. Adoption} & ``\textit{Given that I had access to the system, I predict that I would use it.}'' & \multirow{2}{*}{4.9 (SD=1.50)} & \multirow{2}{*}{6.2 (SD=0.75)} & \multirow{2}{*}{\nnew{$.003^{**}$}} \\
    \addlinespace[.1cm]
    \hline\hline
    \addlinespace[.1cm]
    \multirow{2}{*}{{3. Author Discovery}} & Sum of participants' responses to the 4 questionnaire items below. & \multirow{2}{*}{15.6 (SD=5.76)} & \multirow{2}{*}{21.9 (SD=4.33)} & \multirow{2}{*}{\nnew{$.001^{**}$}} \\
    \midrule
    3a. Finding & ``\textit{The system helped me find relevant authors.}'' & {3.9 (SD=1.82)} & {6.1 (SD=1.34)} & {\nnew{$.002^{**}$}} \\
    \addlinespace[.1cm]
    \multirow{2}{*}{3b. Curiosity} & ``\textit{The system made me curious about authors' research.}'' & \multirow{2}{*}{4.3 (SD=1.58)} & \multirow{2}{*}{6.1 (SD=1.34)} & \multirow{2}{*}{\nnew{$.001^{**}$}} \\
    \addlinespace[.1cm]
    \multirow{2}{*}{3c. Sensemaking} & ``\textit{The system helped me make sense of authors' research.}'' & \multirow{2}{*}{3.5 (SD=1.46)} & \multirow{2}{*}{4.8 (SD=1.53)} & \multirow{2}{*}{\nnew{$.008^{**}$}} \\
    \addlinespace[.1cm]
    {3d. Explanation Helpfulness} & ``\textit{The system’s explanations of relevant authors became more helpful the more I used the system.}'' & \multirow{2}{*}{3.9 (SD=1.77)} & \multirow{2}{*}{4.9 (SD=1.59)} & \multirow{2}{*}{\nnew{$.02^{*}$}} \\
    \bottomrule
    \end{tabular}
    \caption{{Descriptions of additional questionnaire items and responses grouped by condition. $p-$values are from two-sided paired samples \nnew{Wilcoxon's signed rank} tests. The results show that the overall workload was significantly lower in the \sys condition than the \baseline condition. While the adoption plausibility was higher in the \sys condition, the overall TAM responses did not differ significantly between the two conditions. \sys responses were significantly more favorable towards all author discovery helpfulness questions than those of the \baseline condition.}}
    \label{table:full_survey1}
    \vspace{-2em}
\end{table*}

\begin{table*}[t!]
    \centering
    \begin{tabular}{p{2.25cm} p{6.2cm} p{2.0cm} p{2.0cm} p{.85cm}}
    \toprule
    & \textbf{Description} & \textsc{Baseline} & \sys & \textit{p}-val. \\
    \midrule
    \multirow{2}{*}{{4. Paper Discovery}} & Sum of participants' responses to the 3 questionnaire items below. & \multirow{2}{*}{13.6 (SD=4.21)} & \multirow{2}{*}{17.0 (SD=2.85)} & \multirow{2}{*}{\nnew{$.002^{**}$}} \\
    \midrule
    4a. Finding & ``\textit{The system helped me find relevant papers.}'' & {3.9 (SD=1.82)} & {6.1 (SD=1.34)} & {\nnew{$.005^{**}$}} \\
    \addlinespace[.1cm]
    \multirow{2}{*}{4b. Curiosity} & ``\textit{The system made me curious about the papers I found.}'' & \multirow{2}{*}{4.8 (SD=1.64)} & \multirow{2}{*}{6.2 (SD=0.83)} & \multirow{2}{*}{\nnew{$.01^{*}$}} \\
    \addlinespace[.1cm]
    {4c. Explanation Helpfulness} & ``\textit{The system’s explanations of relevant papers became more helpful the more I used the system.}'' & \multirow{2}{*}{3.6 (SD=1.63)} & \multirow{2}{*}{4.6 (SD=1.54)} & \multirow{2}{*}{\nnew{$.052$}} \\
    \addlinespace[.1cm]
    \hline\hline
    \addlinespace[.1cm]
    {{5. Common Features}} & Avg. of participants' responses to the 4 questionnaire items below. & \multirow{2}{*}{3.3 (SD=1.26)} & \multirow{2}{*}{3.5 (SD=1.27)} & \multirow{2}{*}{\nnew{$.25$}} \\
    \midrule
    \multirow{2}{*}{5a. \# of Papers} & ``\textit{I found the authors’ total number of publications useful.}'' (Example provided) & \multirow{2}{*}{3.1 (SD=1.48)} & \multirow{2}{*}{3.8 (SD=1.81)} & \multirow{2}{*}{\nnew{$.07$}} \\
    \addlinespace[.1cm]
    {5b. \# of Relevant Papers} & ``\textit{I found the number of relevant papers estimated by the system useful.}'' (Example provided) & \multirow{2}{*}{3.6 (SD=1.46)} & \multirow{2}{*}{3.8 (SD=1.57)} & \multirow{2}{*}{\nnew{$.65$}} \\
    \addlinespace[.1cm]
    {5c. h-index} & ``\textit{I found authors' h-index useful.}'' & {3.3 (SD=1.39)} & {2.9 (SD=1.39)} & {\nnew{$.15$}} \\
    \addlinespace[.1cm]
    {5d. \# of Citations} & ``\textit{I found authors' citation counts useful.}'' & {3.5 (SD=1.32)} & {3.7 (SD=1.82)} & \nnew{$.72$} \\
    \multirow{2}{*}{5e. Relevance Score} & ``\textit{I found the ``relevance score'' explanation for each paper useful.}'' (Example provided) & \multirow{2}{*}{4.5 (SD=1.90)} & \multirow{2}{*}{4.5 (SD=1.90)} & \multirow{2}{*}{\nnew{$.97$}} \\
    \addlinespace[.1cm]
    \hline\hline
    \addlinespace[.1cm]
    \multirow{2}{*}{{Specific Features}} & Avg. of participants' responses to the condition-specific feature questionnaire items below. & \multirow{2}{*}{4.8 (SD=1.33)} & \multirow{2}{*}{4.8 (SD=0.96)} & \multirow{2}{*}{\nnew{$.74$}} \\
    \addlinespace[.1cm]
    {{Most Favored Feature}} & Avg. of the highest-rated condition specific feature for each participant. & \multirow{2}{*}{5.5 (SD=1.51)} & \multirow{2}{*}{6.2 (SD=0.83)} & \multirow{2}{*}{\nnew{$.05^{*}$}} \\
    \midrule
    \multirow{2}{*}{{Coauthor Filter}} & ``\textit{I found the ``co-authored with [a saved author]'' filter buttons useful.}'' (Example provided) & \multirow{2}{*}{\textsc{na}} & \multirow{2}{*}{5.3 (SD=1.45)} & \multirow{2}{*}{\textsc{na}} \\
    \addlinespace[.1cm]
    \multirow{2}{*}{{Cited-by Filter}} & ``\textit{I found the ``cited by [a saved author]'' filter buttons useful.}'' (Example provided) & \multirow{2}{*}{\textsc{na}} & \multirow{2}{*}{5.3 (SD=1.54)} & \multirow{2}{*}{\textsc{na}} \\
    \addlinespace[.1cm]
    {{Histogram}} & ``\textit{I found the histogram useful.}'' & {\textsc{na}} & {3.8 (SD=1.47)} & {\textsc{na}} \\
    \addlinespace[.1cm]
    \multirow{2}{*}{{Histogram Filter}} & ``\textit{I found being able to see the selected filter counts on the histogram useful.}'' (Example provided) & \multirow{2}{*}{\textsc{na}} & \multirow{2}{*}{4.2 (SD=1.68)} & \multirow{2}{*}{\textsc{na}} \\
    \addlinespace[.1cm]
    {Saved Coauthor Highlight} & ``\textit{I found the ``co-authored with [a saved author]'' explanation for each paper useful.}'' (Example provided) & \multirow{2}{*}{\textsc{na}} & \multirow{2}{*}{4.9 (SD=1.41)} & \multirow{2}{*}{\textsc{na}} \\
    \addlinespace[.1cm]
    {Cited-by Paper Exp.} & ``\textit{I found the ``[a saved author] cited this paper'' explanation for each paper useful.}'' (Example provided) & \multirow{2}{*}{\textsc{na}} & \multirow{2}{*}{5.4 (SD=1.41)} & \multirow{2}{*}{\textsc{na}} \\
    \midrule
    {\textsc{FeedLens}~\cite{kaur-uist22-feedlens} Dots}& ``\textit{I found the green dots next to author names useful.}'' (Example provided) & \multirow{2}{*}{4.8 (SD=1.76)} & \multirow{2}{*}{\textsc{na}} & \multirow{2}{*}{\textsc{na}} \\
    \addlinespace[.1cm]
    \multirow{2}{*}{Tooltip}& ``\textit{I found the author tooltip explanation useful.}'' (Example provided) & \multirow{2}{*}{4.9 (SD=1.45)} & \multirow{2}{*}{\textsc{na}} & \multirow{2}{*}{\textsc{na}} \\
    \bottomrule
    \end{tabular}
    \caption{{Descriptions of additional questionnaire items and responses grouped by condition, continued. $p-$values are from two-sided paired samples \nnew{Wilcoxon's signed rank} tests. \sys responses were significantly more favorable towards all paper discovery helpfulness questions than those of the \baseline condition. For common features responses did not show any significant difference between the two conditions. Furthermore, the average responses to system-specific feature questions showed no significant advantage of one system over the other. However, aggregating over the most favored feature from each user, \sys showed a significantly higher average.}}
    \label{table:full_survey2}
    \vspace{-2em}
\end{table*}

\end{document}